\documentclass[12pt]{article} 

\usepackage{amsmath,amssymb,amsfonts}
\usepackage{psfrag}
\usepackage{color}
\definecolor{darkblue}{rgb}{0.1,0.1,.7}
\usepackage[colorlinks, linkcolor=darkblue, citecolor=darkblue, urlcolor=darkblue, linktocpage]{hyperref} 
\usepackage[square, comma, compress,numbers]{natbib}
\usepackage[]{amsmath}
\usepackage[]{graphicx}
\usepackage[]{latexsym}
\usepackage{geometry}
\usepackage{amscd}
\usepackage[all,cmtip]{xy}
\usepackage{mathrsfs}
\usepackage[margin=10pt,font=small,labelfont=bf]{caption}
\usepackage{dsdshorthand}
\usepackage{simplewick}
\usepackage{changepage}
\numberwithin{equation}{section}

\renewcommand{\be}{\begin{eqnarray}}
\renewcommand{\ee}{\end{eqnarray}}
\newcommand{\bea}{\begin{eqnarray}}
\newcommand{\eea}{\end{eqnarray}}

\def\beq{\begin{equation}} 
\def\eeq{\end{equation}} 
\def\<{\langle}
\def\>{\rangle}

\def\nn{\nonumber} 
 
\def\cO {{\cal O}} 
\def\cG {{\cal G}} 
 
\def\cN {{\cal N}}

\newcommand{\cX}{\mathcal{X}}
\newcommand{\cS}{\mathcal{S}}
\newcommand{\cY}{\mathcal{Y}}
\newcommand{\cJ}{\mathcal{J}}

\def\CO{{\cal O}}

\begin{document}

\vspace*{-.6in} \thispagestyle{empty}
\begin{flushright}
\end{flushright}
\vspace{.2in} {\Large
\begin{center}
{ $\mathcal{N}=1$ Superconformal Blocks for General Scalar Operators\\\vspace{.1in}}
\end{center}
}
\vspace{.2in}
\begin{center}
{\bf 
Zuhair U. Khandker$^{a}$,
Daliang Li$^{b}$, 
David Poland$^{b}$,
David Simmons-Duffin$^{c}$} 
\\
\vspace{.2in} 
$^a$ {\it  Physics Department, Boston University, Boston, MA 02215}\\
$^b$ {\it  Department of Physics, Yale University, New Haven, CT 06511}\\
$^c$ {\it School of Natural Sciences, Institute for Advanced Study, Princeton, New Jersey 08540}
\end{center}

\vspace{.2in}

\begin{abstract}
We use supershadow methods to derive new expressions for superconformal blocks in 4d $\cN=1$ superconformal field theories. We analyze the four-point function $\<\cA_1 \cA_2^\dag \cB_1 \cB_2^\dag\>$, where $\cA_i$ and $\cB_i$ are scalar superconformal primary operators with arbitrary dimension and $R$-charge and the exchanged operator is neutral under $R$-symmetry. Previously studied superconformal blocks for chiral operators and conserved currents are special cases of our general results.
 
\end{abstract}

\newpage

\tableofcontents

\newpage


\section{Introduction}
\label{sec:intro}

Significant progress has been made recently in the conformal bootstrap program~\cite{Polyakov:1974gs} for theories in higher than two spacetime dimensions~\cite{Rattazzi:2008pe,Rychkov:2009ij,Caracciolo:2009bx,Poland:2010wg,Rattazzi:2010gj,Rattazzi:2010yc,Vichi:2011ux,Poland:2011ey,Rychkov:2011et,ElShowk:2012ht,Liendo:2012hy,ElShowk:2012hu,Fitzpatrick:2012yx,Komargodski:2012ek,Beem:2013qxa,Kos:2013tga,Gliozzi:2013ysa,El-Showk:2013nia,Alday:2013opa,Gaiotto:2013nva,Bashkirov:2013vya,Beem:2013sza,Berkooz:2014yda,El-Showk:2014dwa,Gliozzi:2014jsa,Fitzpatrick:2014vua,Nakayama:2014lva,Beem:2014kka}. In particular, spectacular results have emerged from applying the bootstrap to supersymmetric systems~\cite{Poland:2010wg,Vichi:2011ux,Poland:2011ey,Beem:2013qxa,Bashkirov:2013vya,Beem:2013sza,Berkooz:2014yda,Beem:2014kka}, where constraints from supersymmetry and knowledge of protected aspects of the spectrum make the approach even more powerful. A crucial ingredient in the superconformal bootstrap is the expansion of four-point functions in superconformal blocks, which sum up the contributions of all of the descendants of a given superconformal primary operator. Results for 4d superconformal blocks in $\cN=1,2,4$ theories have previously appeared in~\cite{Poland:2010wg,Dolan:2001tt,Dolan:2004iy,Dolan:2004mu,Fortin:2011nq,Berkooz:2014yda}. 

Recently, we introduced a new covariant approach to studying superconformal blocks~\cite{Fitzpatrick:2014oza}, based on generalizing the shadow formalism developed in~\cite{Ferrara:1972xe,Ferrara:1972ay,Ferrara:1972uq,Ferrara:1973vz,SimmonsDuffin:2012uy} to superconformal theories. In~\cite{Fitzpatrick:2014oza} we used this approach to analyze four-point functions of chiral and antichiral operators in theories with $\cN=1,2$ superconformal symmetry. In the present work we will apply our formalism to four-point functions containing general scalar operators in $\cN=1$ theories, focusing on situations where the exchanged operator is neutral under the $U(1)_R$ symmetry.

The class of correlators we consider includes the interesting cases of chiral-antichiral four-point functions, for which the bootstrap was performed in \cite{Poland:2010wg,Vichi:2011ux,Poland:2011ey}, and also four-point functions of currents, which have been studied in \cite{Fortin:2011nq,Berkooz:2014yda}. These types of correlators (together with mixed chiral-current correlators which are also covered by our formalism) are extremely fruitful objects of study in the superconformal bootstrap for three reasons.  Firstly, we have extensive knowledge of the protected spectrum of $\cN=1$ superconformal theories.  Secondly, four-point functions of scalars are currently the easiest systems for applying numerical bootstrap techniques.  Thirdly,  bootstrap techniques are often most powerful for four-point functions of low-dimension operators, and such operators are often protected.  For these reasons, our expressions will likely be crucial ingredients in future explorations of the 4d $\cN=1$ bootstrap.

The initial complication that arises in our analysis is the fact that multiple structures can appear in superspace three-point functions. Thus, our first task is to review the superembedding formalism for describing these structures and then to enumerate them, which we do in Sections~\ref{sec:superembedding} and~\ref{sec:correlation}. In Section~\ref{sec:PartialWave} we set up and evaluate the superconformal integrals relevant for computing superconformal blocks, with our results given in Section~\ref{sec:results}. We also show how the cases of four-point functions containing chiral or conserved current operators emerge as special cases of our general result. In Section~\ref{sec:decomp} we show explicitly how previous results for $\cN=2$ superconformal blocks decompose into $\cN=1$ superconformal blocks, providing a highly nontrivial consistency check on the form of the blocks. Several details of our calculations are presented in the appendices. 

\section{The Superembedding and Supershadow Formalisms}
\label{sec:superembedding}

\subsection{Superembedding Space}
\label{sec:superembeddingspace}

The superembedding formalism provides a simple language for writing down and classifying superconformally invariant correlation functions in $\cN=1$ SCFTs \cite{Goldberger:2011yp,Goldberger:2012xb,Khandker:2012pa,Siegel:1992ic,Siegel:1994cc,Siegel:2010yd,Siegel:2012di,Kuzenko:2006mv,Kuzenko:2012tb,Maio:2012tx}.  The essential idea is the one underlying the embedding formalism \cite{Dirac:1936fq, Mack:1969rr, Boulware:1970ty, Ferrara:1973eg, Weinberg:2010fx,Costa:2011mg,Costa:2011dw,SimmonsDuffin:2012uy}.  We introduce a space on which the superconformal group $\SU(2,2|1)$ acts linearly, and view SCFT operators as functions on this space (with special properties depending on the operator's dimension, spin, and $R$-charge).  Correlators are then given by products of simple invariants. This story and associated techniques for computing superconformal blocks were developed recently in~\cite{Fitzpatrick:2014oza}; here we will briefly summarize the results we need for our computation.

The basic superconformally covariant objects are supertwistors
\be
\cZ_A =
\begin{pmatrix}
Z_\a\\
Z^{\dot\a}\\
Z_5
\end{pmatrix}\in \C^{4|1},
\ee
and dual supertwistors
\be
\bar \cZ^A =
\begin{pmatrix}
\bar Z^\a &
\bar Z_{\dot\a} &
\bar Z^5
\end{pmatrix}\in \C^{4|1}.
\ee
They transform as fundamentals and antifundamentals of $\SU(2,2|1)$, so that the pairing $\bar \cZ^A_1 \cZ_{2A}$ is $\SU(2,2|1)$-invariant.

Superspace is given by a pair of supertwistors $\cZ_A^a$, $a=1,2$, and a pair of dual supertwistors $\bar \cZ^{\dot a A}$, $\dot a=1,2$, subject to a constraint
\be
\label{eq:supertwistorconstraint}
\bar \cZ^{\dot a A}\cZ_A^a = 0,\qquad a,\dot a = 1,2
\ee
and with gauge redundancies
\be
\label{eq:gaugeredundancies}
\cZ^{a}_A\sim \cZ^b_A g_b{}^a, \qquad \bar \cZ^{\dot a A}\sim \bar g^{\dot a}{}_{\dot b}\bar \cZ^{\dot b A},\qquad\textrm{for}\qquad g,\bar g\in \GL(2,\C).
\ee
Here, ``$\sim$" means ``is equivalent to."  This space has a natural action of the superconformal group given by matrix multiplication on the $\SU(2,2|1)$ indices $A$.  On the other hand, it is equivalent to the usual $\cN=1$ superspace.  To see why, one can choose the ``Poincar\'e section" gauge fixing of $\GL(2,\C)\x\GL(2,\C)$, where $(\cZ,\bar \cZ)$ take the form
\be
\cZ_A^a = \begin{pmatrix}
\de_\a{}^a\\
ix_+^{\dot\a a}\\
2\th^a
\end{pmatrix},
\qquad
\bar \cZ^{\dot a A} = \begin{pmatrix}
-ix_-^{\dot a \a} & \de^{\dot a}_{\dot \a} & 2\bar\th^{\dot a}
\end{pmatrix}.
\label{eq:poincareslice}
\ee
The constraint (\ref{eq:supertwistorconstraint}) then reads $x_+-x_--4i\bar\th\th=0$, so that we can identify $x_{\pm}$ with the usual chiral/anti-chiral bosonic coordinates and $\th,\bar\th$ with the usual fermionic coordinates on superspace.  Any function of $\cZ$'s ($\bar \cZ$'s) alone is purely chiral (anti-chiral).

We will often work with bi-supertwistors
\be
\cX_{AB} \equiv \cZ^a_A\cZ^b_B\e_{ab},\qquad \bar \cX^{AB}\equiv \bar \cZ^{\dot a A}\bar \cZ^{\dot b B}\e_{\dot a \dot b}
\ee
which are invariant under the $\SL(2,\C)\x\SL(2,\C)$ subgroup of the gauge redundancies (\ref{eq:gaugeredundancies}).  A basic set of superconformal invariants are given by supertraces of products of $\cX$'s and $\bar \cX$'s, for instance
\be
\< \bar{2}1\> &\equiv& \bar{\cX}_{2}^{AB}\cX_{1BA},\\
\<\bar 4 3 \bar 2 1\>   &\equiv& \bar{\cX}_4^{AB}\cX_{3BC}\bar \cX_2^{CD}\cX_{1DA} (-1)^{p_C}.
\ee
Here, $p_C$ denotes the fermion number parity of the index $C$ ($1$ if $C=5$, and $0$ otherwise).\footnote{The rule for inserting signs $(-1)^{p_A}$ into products of supermatrices is that we need a sign whenever superindices $A,B$ are contracted from bottom to top, since the basic superconformally invariant pairing contracts indices from top to bottom.}
By construction, these invariants are chiral in unbarred coordinates and anti-chiral in barred coordinates.

\subsection{Lifting $\cN=1$ Fields to Superembedding Space}
\label{subsec:lifting}

A four-dimensional $\mathcal{N}=1$ superconformal primary is labeled by its $\SL(2,\mathbb{C})$ Lorentz quantum numbers $(\frac{j}{2},\frac{\bar{j}}{2})$, its scaling dimension $\Delta$, and its $U(1)_R$ charge $R$. It is convenient to summarize these labels as $(\frac{j}{2},\frac{\bar{j}}{2},q,\bar{q})$, where the {\it superconformal weights} $q,\bar{q}$ are given by
\begin{equation}
q \equiv \frac{1}{2} \left(\Delta + \frac{3}{2}R\right), \hspace{5mm} \bar{q} \equiv \frac{1}{2} \left(\Delta - \frac{3}{2}R\right) .
\label{q}
\end{equation}

A general superfield with spin lifts to a multi-twistor operator in superembedding space
\be
\phi_{\alpha_1\cdots\alpha_j}^{\dot{\beta}_1\cdots\dot{\beta}_{\bar{j}}}
&\to&
\Phi_{B_1\cdots B_{\bar{j}}}^{\phantom{A_{j}\cdots A_{1}}A_{1}\cdots A_{j}}(\cX,\bar \cX),
\ee
with homogeneity determined by its superconformal weights
\be
\Phi(\l \cX,\bar \l \cX) = \l^{-(2q+j)/2}\bar\l^{-(2\bar q+ \bar j)/2}\Phi(\cX,\bar \cX).
\ee
The field $\Phi$ is also subject to gauge redundancies in each index,
\be
\Phi_{B_1\cdots B_{\bar{j}}}^{\phantom{A_{j}\cdots A_{1}}A_{1}\cdots A_{j}}(\cX,\bar \cX) &\sim& \Phi_{B_1\cdots B_{\bar{j}}}^{\phantom{A_{j}\cdots A_{1}}A_{1}\cdots A_{j}}(\cX,\bar \cX)+\cX_{B_1C}\Lambda_{B_2\cdots B_{\bar j}}{}^{C A_1\cdots A_j},
\ee
and similarly for the other indices.  It is convenient to introduce index-free notation by using auxiliary twistors $\cS_{A},\bar{\cS}^{A}$ to absorb the indices of the superembedding fields. We define
\begin{equation}
\Phi(\cX,\bar{\cX},\cS,\bar{\cS})\equiv\bar{\cS}^{B_{\bar{j}}}\cdots\bar{\cS}^{B_{1}}\Phi_{B_{1}\cdots B_{\bar{j}}}^{\phantom{A_{n}\cdots A_{1}}A_{1}\cdots A_{j}}\cS_{A_{j}}\cdots \cS_{A_{1}} .
\label{IndexFreeField}
\end{equation}
The gauge-redundancy of $\Phi$ allows us to restrict $\cS,\bar{\cS}$ to be transverse and null\footnote{Nullness follows because the transverse conditions can be solved by $\cS=\cX\bar{\cT}$, $\bar{\cS}=\bar{\cX}\cT$ for some $\cT,\bar{\cT}$.}
\begin{equation}
\bar{\cX}\cS=0,\qquad\qquad\bar{\cS}\cX=0,\qquad\qquad\bar{\cS}\cS=0.
\label{transverse}
\end{equation} 
Finally, the four-dimensional superfield is recovered by 
\begin{equation}
\phi_{\alpha_1\cdots\alpha_j}^{\dot{\beta}_1\cdots\dot{\beta}_{\bar{j}}} = \left.\frac{1}{j!}\frac{1}{\bar{j}!} \left(\bar{\cX}\overrightarrow{\partial_{\bar{\cS}}}\right)^{\dot{\beta}_1}\cdots\left(\bar{\cX}\overrightarrow{\partial_{\bar{\cS}}}\right)^{\dot{\beta}_{\bar{j}}} \Phi(\cX,\bar{\cX},\cS,\bar{\cS}) \left(\overleftarrow{\partial_\cS} \cX\right)_{\alpha_1} \cdots \left(\overleftarrow{\partial_\cS} \cX\right)_{\alpha_j} \right|_\textrm{Poincar\'e} .
\label{GeneralProj}
\end{equation}
where the subscript ``Poincar\'e" means we choose the Poincar\'e section gauge fixing (\ref{eq:poincareslice}).

\subsection{Superconformal Integration}

The superspace defined in Section~\ref{sec:superembeddingspace} admits a natural notion of superconformally invariant integration.  Note that the measure
\be
\w &\equiv& \prod_{a=1,2} d^{4|1}\cZ^a\prod_{\dot a=1,2} d^{4|1}\bar \cZ^{\dot a} \de^4(\bar \cZ^{\dot b A}\cZ^b_{A})
\ee
is superconformally invariant, and because of the delta function it is supported on $\cZ,\bar\cZ$ which satisfy the constraint (\ref{eq:supertwistorconstraint}).   The form $\w$ transforms in the following way under the gauge redundancies~(\ref{eq:gaugeredundancies}):
\be
\w
&\to&
(\det g) (\det \bar g) \w.
\ee
Suppose $f(\cZ,\bar \cZ)$ is a function that transforms oppositely under $\GL(2,\C)\x\GL(2,\C)$:
\be
f(\cZ g, \bar g \bar \cZ) &=& (\det g)^{-1} (\det \bar g)^{-1} f(\cZ,\bar \cZ).
\label{eq:integrandtransformation}
\ee
Then the product $\w f(\cZ,\bar \cZ)$ is gauge-invariant and can be integrated, provided we divide by the volume of the gauge group
\be
\label{eq:superconformalintegral}
\int D[\cZ,\bar \cZ] f(\cZ,\bar \cZ) &\equiv& \frac{1}{\mathrm{vol}(\GL(2,\C)\x\GL(2,\C))}\int \omega f(\cZ,\bar \cZ).
\ee
This gauge-redundant integral is defined by the Faddeev-Popov procedure: we choose a gauge slice for $\GL(2,\C)\x\GL(2,\C)$, introduce the appropriate determinant, and integrate over the remaining variables.

Note that any $f$ satisfying the condition~(\ref{eq:integrandtransformation}) can always be written as a homogeneous function of $\cX,\bar \cX$ with degree $-1$ in both variables, so we will sometimes write $f(\cX,\bar \cX)$ instead of $f(\cZ,\bar \cZ)$.  We will also occasionally write $D[\cX,\bar \cX]$ for $D[\cZ,\bar \cZ]$.

A special class of superconformal integrals will be particularly important in our computations.  This is the case where $f(\cX,\bar \cX)$ is independent of the fermionic variables 
\be
\eta^a\equiv Z_5^a,\qquad\bar \eta^{\dot a}\equiv \bar Z^{\dot a5},
\ee
so we may write $f(\cX,\bar \cX)=f(X,\bar X)$, where $X_{\s\rho}$ and $\bar X^{\s\rho}$ are the restrictions of $\cX_{AB}$ and $\bar \cX^{AB}$ to bosonic twistor indices $\s,\rho=1,2,3,4$.  The $4\x4$ antisymmetric matrices $X,\bar X$ can also be thought of as six-dimensional vectors $X,\bar X\in \C^6$ transforming under $\SO(4,2)\cong \SU(2,2)$.\footnote{Our conventions are: $(\tl{\Gamma}^{m}\Gamma^{n}+\tl{\Gamma}^{n}\Gamma^{m})_{\alpha}^{\phantom{\alpha}\beta}=-2\eta^{mn}\delta_{\alpha}^{\phantom{\alpha}\beta}$, $X^{\alpha\beta}=\frac{1}{2}X_{m}\Gamma^{m\alpha\beta}$, $X_{\alpha\beta}=\frac{1}{2}X_{m}\tl{\Gamma}_{\alpha\beta}^{m}$.}
In~\cite{Fitzpatrick:2014oza}, we showed that such superconformal integrals can be computed in a simple way in terms of non-supersymmetric conformal integrals,
\be
\label{eq:specialintegralformula}
\int D[\cX,\bar \cX]f(X,\bar X)=\int D^4 X \left.\ptl_{\bar X}^2 f(X,\bar X)\right|_{\bar X=X}.
\ee
Here, the notation $\ptl_{\bar X}^2$ means we take two derivatives with respect to $\bar X$ as an independent variable, and contract indices using the six-dimensional metric.\footnote{In $\SU(2,2)$ notation, this is $\ptl_{\bar X}^2\propto\e^{\a\b\g\de}\ptl_{\bar X^{\a\b}}\ptl_{\bar X^{\g\de}}$.}  The conformally-invariant
measure $D^4X$ on the right-hand side of Eq.~(\ref{eq:specialintegralformula}) was defined in~\cite{SimmonsDuffin:2012uy} and is given by\footnote{As written, the measures $D[\cX,\bar\cX]$ and $D^4 X$ are ambiguous under multiplication by an overall constant.  This is because division by the infinite volumes $\mathrm{vol}(\GL(2,\C)\x\GL(2,\C))$ and $\mathrm{vol}(\GL(1,\C))$ is really {\it defined} by a choice of Faddeev-Popov determinant in the gauge-fixing procedure, and the overall scale of this determinant is arbitrary.  In this work, we choose determinants so that~(\ref{eq:specialintegralformula}) holds.  In other words, we absorb any difference in normalization of the two sides into the definition of $D[\cX,\bar\cX]$.  The overall normalization of $D^4 X$ will drop out of our computations and is unimportant (in practice we choose the same normalization as in \cite{SimmonsDuffin:2012uy}).} 
\be
\int D^{4}Xf(X)\equiv\frac{1}{\mathrm{vol}(\GL_{1})}\int d^{6}X\delta(X^{2})f(X).\label{Conf Measure}
\ee

In this work, we will also encounter more general $\cN=1$ superconformal integrals, where $f(\cX,\bar\cX)$ is not independent of $\eta,\bar\eta$ and~(\ref{eq:specialintegralformula}) doesn't immediately apply.  For our purposes, we will be able to fix the required answers without going through a full computation.  But for now, let us note how these integrals can be done in principle.

Consider a general function $f(\cX,\bar \cX)$ and expand in the fermionic variables $\eta,\bar \eta$.  The measure $\w$ contains a delta function
\be
\de^4(\bar\cZ^{\dot a A}\cZ^a_A)&=&\de^4(\bar Z^{\dot a}\.Z^a+\bar\eta^{\dot a}\eta^a),
\ee
where $\bar Z\.Z$ is the $\SU(2,2)$-invariant pairing between the bosonic components of $\bar \cZ,\cZ$.
The presence of this delta function means that in any term with equal degree in $\eta,\bar\eta$, we can replace 
\be
\bar\eta^{\dot a}\eta^a &\to& -\bar Z^{\dot a}\.Z^a,
\ee
leaving the integral unchanged.  Meanwhile, terms with unequal degree in $\eta,\bar\eta$ integrate to zero.  Thus, via the above replacement we can completely remove the $\eta,\bar\eta$ dependence of $f(\cX,\bar\cX)$ and reduce to the case where~(\ref{eq:specialintegralformula}) applies.

\subsection{Supershadows}
\label{sec:supershadows}

For an operator $\mathcal{O}(\cX,\bar{\cX},\cS,\bar{\cS})$ with quantum numbers $( \frac{j}{2},\frac{\bar{j}}{2},q,\bar{q})$, we define the nonlocal {\it shadow operator}
\begin{equation}
\tl{\mathcal{O}}(\cX,\bar{\cX},\cS,\bar{\cS})\equiv\int D[\cY,\bar{\cY}]\frac{1}{\langle \cX\bar{\cY}\rangle ^{1-q+\frac{j}{2}}\langle \bar{\cX}\cY\rangle ^{1-\bar{q}+\frac{\bar{j}}{2}}}{\mathcal{O}^\dag}(\cY,\bar\cY,\cY\bar{\cS},\bar{\cY}\cS),
\label{Shadow}
\end{equation}
where $D[\cY,\bar{\cY}]$ is the superconformal measure from Eq.~(\ref{eq:superconformalintegral}) and ${\mathcal{O}^\dag}\sim ( \frac{\bar{j}}{2},\frac{j}{2})$ transforms in the conjugate Lorentz representation to $\mathcal{O}$.  Because the integrand and measure are superconformally covariant, $\tl \cO$ transforms like a superconformal primary with quantum numbers $(\frac{\bar{j}}{2},\frac{j}{2},1-q,1-\bar{q})$.  The shadow transform~(\ref{Shadow}) is uniquely determined up to a constant by the requirement that the integrand transform appropriately under $\GL(2,\C)\x\GL(2,\C)$~(\ref{eq:integrandtransformation}), together with the transverseness conditions on auxiliary twistors~(\ref{transverse}).

Within a correlation function, one can project onto the superconformal multiplet of $\mathcal{O}$ by inserting the projector
\begin{equation}
\left|\mathcal{O}\right|= \left. \frac{1}{j!^{2}\bar{j}!^{2}}\int D[\cX,\bar{\cX}]|\mathcal{O}(\cX,\bar{\cX},\cS,\bar{\cS})\rangle \left(\overleftarrow{\partial_{\cS}}\cX\overrightarrow{\partial_{\cT}}\right)^{j}\left(\overleftarrow{\partial_{\bar{\cS}}}\bar{\cX}\overrightarrow{\partial_{\bar{\cT}}}\right)^{\bar{j}}\langle \tl{\mathcal{O}}(\cX,\bar{\cX},\cT,\bar{\cT})| \hspace{2mm} \right|_{M}
\label{Projector}
\end{equation}
In the definition above,
$\left.\right|_{M}$ schematically denotes a ``monodromy projection''~\cite{SimmonsDuffin:2012uy}.%
\footnote{\noindent As detailed in~\cite{SimmonsDuffin:2012uy}, the monodromy projection restricts
the region of integration in Eq.~(\ref{Projector}) so as not to interfere
with the OPE expansion of the fields in a four-point function, thus
avoiding extraneous ``shadow'' contributions in the computation
of partial waves.%
} Operationally, all we will need is the result for a monodromy-projected
(non-supersymmetric) conformal integral, which has been derived previously~\cite{DO1,SimmonsDuffin:2012uy} and is given in Eq.~(\ref{ConfInt}).  The form of (\ref{Projector}) is uniquely determined by superconformal invariance and the various conditions on $\cO$ as a function of $\cX,\cS,\cT$ and their conjugates.  

The example of interest for us will be a four-point function $\< \cA_1 \cA_2^\dag \cB_1 \cB_2^\dag\>$, where the superconformal partial wave $\mathcal{W}_{\mathcal{O}}$ corresponding to $\mathcal{O}\in \cA_1\x\cA_2^\dag$ is given (up to some normalization) by
\begin{equation}
\mathcal{W_{O}}\propto  \< \cA_{1}\cA_{2}^\dag \left|\mathcal{O}\right| \cB_{1}\cB_{2}^\dag \>.
\label{PWave}
\end{equation}
The partial wave $\mathcal{W_O}$ differs from the superconformal block $\cG_\cO$ by simple kinematic factors.
In general the three-point functions $\< \cA_1 \cA_2^\dag \CO \>$ and $\< \tl{\cO} \cB_1 \cB_2^\dag \>$ appearing in $\mathcal{W_{O}}$ contain multiple structures, each with an independent coefficient. As we show explicitly below, $\mathcal{W_{O}}$ will then receive independent contributions from each of these structures.

\section{Correlation Functions}
\label{sec:correlation}

\subsection{2-Point Functions}

The two-point correlation function of a scalar superfield with its conjugate
is determined by superconformal symmetry up to a constant. This fact
is obvious in the superembedding space where there is only a single
superconformal invariant with the correct homogeneity in $\cX_{1,2},\bar\cX_{1,2}$,
\begin{equation}
\langle \mathcal{A}(1,\bar{1})\mathcal{A}^{\dagger}(2,\bar{2})\rangle \propto\frac{1}{\langle 1\bar{2}\rangle ^{q_{\cA}}\langle 2\bar{1}\rangle ^{\bar{q}_{\cA}}}\label{eq:A2-PF}
\end{equation}
where we have labeled the coordinates $(\cX_{i},\bar{\cX}_{i})$ simply as $(i,\bar{i})$. 
After restricting to the Poincare section, $\< \bar{i}j \>$ reduces to $\frac{1}{2}x_{\bar{i}j}^2$, where $x_{\bar{i}j}^\mu \equiv x_{i-}^\mu - x_{j+}^\mu +2i\theta_j\sigma^\mu\bar{\theta}_i$, giving
\begin{equation}
\langle \mathcal{A}(z_{1})\mathcal{A}^\dag(z_{2})\rangle \propto\frac{1}{x_{\bar{2}1}^{2q_\cA}x_{\bar{1}2}^{2\bar q_\cA}}\label{eq:A2PF4D}
\end{equation}
where $z_{i}\equiv(x_{i},\theta_{i},\bar{\theta}_{i})$. 

Correlation functions of operators with spin are most easily expressed
using index free notation. The two-point function between an $\mathcal{N}=1$
superfield and its conjugate is determined by superconformal symmetry
up to a constant as
\begin{equation}
\langle \mathcal{O}(1,\bar{1},\cS_{1},\bar{\cS}_{1})\mathcal{O}^\dag(2,\bar{2},\cS_{2},\bar{\cS}_{2})\rangle \propto\frac{(\cS_{1}\bar{\cS}_{2})^{j}(\cS_{2}\bar{\cS}_{1})^{\bar{j}}}{\langle 1\bar{2}\rangle ^{q+\frac{j}{2}}\langle 2\bar{1}\rangle ^{\bar{q}+\frac{\bar{j}}{2}}}.\label{eq:O2PF}
\end{equation}
This reproduces the superconformal two-point function when we project to 4d: 
\begin{equation}
\langle \mathcal{O}^{\dot{\beta}_{1}\dots\dot{\beta}_{\bar{j}}}_{\alpha_{1}\dots\alpha_{j}}(z_1){\mathcal{O}^{\dagger}}^{\dot{\alpha}_{1}\dots\dot{\alpha}_{j}}_{\beta_{1}\dots\beta_{\bar{j}}}(z_2)\rangle \propto
\frac{C_{(\alpha_1\cdots\alpha_j)  (\beta_1\cdots \beta_{\bar{j}}) }^{(\dot{\alpha}_1\cdots\dot{\alpha}_j)(\dot{\beta}_1\cdots\dot{\beta}_{\bar{j}})}}{\left(x_{\bar{2}1}^{2}\right)^{q+\frac{j}{2}}\left(x_{\bar{1}2}^{2}\right)^{\bar{q}+\frac{\bar{j}}{2}}},\nn
\end{equation}
\begin{equation}
C_{\alpha_1\cdots\alpha_j \beta_1\cdots \beta_{\bar{j}}}^{\dot{\alpha}_1\cdots\dot{\alpha}_j\dot{\beta}_1\cdots\dot{\beta}_{\bar{j}}} \equiv (x_{\bar{2}1})_{\alpha_1}^{\dot{\alpha}_1} \ldots (x_{\bar{2}1})_{\alpha_j}^{\dot{\alpha}_j}(x_{\bar{1}2})_{\beta_1}^{\dot{\beta}_1}\ldots (x_{\bar{1}2})_{\beta_{\bar{j}}}^{\dot{\beta}_{\bar{j}}}.
\label{eq:O2PF4D} 
\end{equation}

\subsection{3-Point Functions}

In this section we construct the superfield three-point correlation function
$\langle \mathcal{A}_{1}\mathcal{A}_{2}^{\dagger}\mathcal{O}\rangle $,
where $\mathcal{A}_{1,2}$ are scalars with identical
superconformal weights $(q_{\cA},\bar{q}_{\cA})$, and $\mathcal{O}\sim(\frac{\ell}{2},\frac{\ell}{2},\frac{\Delta}{2},\frac{\Delta}{2})$
is a real superfield with dimension $\Delta$ and spin $\ell$. Since
superconformal invariance is explicit in superembedding space, a correlator
is simply the most general function of independent invariants and
tensor structures, consistent with the homogeneity properties of the
participating operators. We will show that this correlation function contains
4 independent coefficients. In special cases such as $\mathcal{A}_{1,2}$
being the chiral or conserved current superfields, there are additional
constraints on the three-point function coefficients. 

The general methods for constructing invariants and tensors in
superembedding space are detailed in~\cite{Goldberger:2012xb,Fitzpatrick:2014oza}. Here we show the
relevant results. The three-point function $\langle \mathcal{A}_{1}\mathcal{A}_{2}^{\dagger}\mathcal{O}(0,\bar{0}) \rangle $
 depends on 2-traces $\langle i\bar{j}\rangle $, where
$i,j\in\{1,2,0\}$. From these 2-traces we can construct an invariant
cross-ratio: 
\begin{equation}
z=\frac{\langle 2\bar{1}\rangle \langle 0\bar{2}\rangle \langle 1\bar{0}\rangle -\langle 1\bar{2}\rangle \langle 2\bar{0}\rangle \langle 0\bar{1}\rangle }{\langle 2\bar{1}\rangle \langle 0\bar{2}\rangle \langle 1\bar{0}\rangle +\langle 1\bar{2}\rangle \langle 2\bar{0}\rangle \langle 0\bar{1}\rangle }.\label{eq:z}
\end{equation}
This cross-ratio vanishes in the limit $\theta_{i}=\bar{\theta}_{i}=0,\ \ i\in\{1,2,0\}$,
where $\langle i\bar{j}\rangle =\langle j\bar{i}\rangle $.
It also has the following properties: 
\begin{equation}
z^{3}=0,\ \ \ z|_{1\leftrightarrow2}=-z.\label{eq:zProperty}
\end{equation}
In addition, the correlator can depend on two elementary tensor structures
\begin{equation}
S\equiv\frac{\bar{\cS}1\bar{2}\cS}{\langle 1\bar{2}\rangle },\ \ \ S|_{1\leftrightarrow2}\equiv\frac{\bar{\cS}2\bar{1}\cS}{\langle 2\bar{1}\rangle }.\label{eq:TensorS}
\end{equation}
or equivalently
\begin{equation}
S_{\pm}=\frac{1}{2}(S\pm(1\leftrightarrow2)),\label{eq:TensorSpm}
\end{equation}
where $\bar{\cS}1\bar{2}\cS$ denotes $\bar{\cS}\cX_{1}\bar{\cX}_{2}\cS$. The
structure $S_{+}$ is nilpotent and satisfies the following relations:
\begin{equation}
\left.S_{+}\right|_{\theta_{i}=\bar{\theta}_{i}=0}=0,\ \ \ (S_{+})^{2}=0,\ \ \ zS_{+}=\frac{1}{2}z^2 S_{-}.\label{eq:SpProperty}
\end{equation}
There are thus two independent spin-$\ell$ tensor structures: 
\begin{eqnarray}
S_{-}^{\ell} & = & \frac{1}{2}\left(S^{\ell}+(-1)^{\ell}(1\leftrightarrow2)\right),\nonumber \\
S_{-}^{\ell-1}S_{+} & = & \frac{1}{2\ell}\left(S^{\ell}-(-1)^{\ell}(1\leftrightarrow2)\right).\label{eq:3PFBasisChange}
\end{eqnarray}

\subsubsection{General 3-Point Functions}

The three-point correlation function $\langle \mathcal{A}_{1}\mathcal{A}_{2}^{\dagger}\mathcal{O}\rangle$
in general contains four independent structures and can be written in the following form:
\begin{equation}
\langle \mathcal{A}_{1}(1,\bar{1})\mathcal{A}_{2}^{\dagger}(2,\bar{2})\mathcal{O}(0,\bar{0},\cS,\bar{\cS})\rangle =\frac{\left(\lambda_{\cA_1\cA_2^\dag\cO}^{(0)}+\lambda_{\cA_1\cA_2^\dag\cO}^{(1)}z+\lambda_{\cA_1\cA_2^\dag\cO}^{(2)}z^{2}\right)S_{-}^{\ell}+\lambda_{\cA_1\cA_2^\dag\cO}^{(3)}S_{+}S_{-}^{\ell-1}}{\langle 1\bar{2}\rangle ^{q_{\cA}-\frac{1}{4}(\Delta+\ell)}\langle 2\bar{1}\rangle ^{\bar{q}_{\cA}-\frac{1}{4}(\Delta+\ell)}(\langle 0\bar{1}\rangle \langle 1\bar{0}\rangle \langle 0\bar{2}\rangle \langle 2\bar{0}\rangle)^{\frac{1}{4}(\Delta+\ell)}},\label{eq:AAO}
\end{equation}
where $\l_{\cA_1\cA_2^\dag\cO}^{(i)}$ are constant coefficients.
The uniqueness of these structures follows from the relations~(\ref{eq:zProperty}, \ref{eq:SpProperty}) above.\footnote{The enumeration of structures can also be done straightforwardly using the formalism of~\cite{Osborn:1998qu}, where we can identify the invariants $J = 2z$ and $I = 2z^2$.}   The $\lambda^{(i)}_{\cA_1\cA_2^\dag\cO}$ are generically unrelated. However, if we impose shortening constraints on $\mathcal{A}_{1,2}$, such as making them chiral or conserved current multiplets, then there will be relations among the $\lambda^{(i)}_{\cA_1\cA_2^\dag\cO}$ coefficients.

\subsubsection{Chiral Operators}

In previous applications of the superembedding space and shadow formalisms, we've written down the chiral three-point function: 
\begin{equation}
\langle \Phi(1,\bar{1})\Phi^{\dagger}(2,\bar{2})\mathcal{O}(0,\bar{0},\cS,\bar{\cS})\rangle =\lambda_{\Phi\Phi^{\dagger}\mathcal{O}}\frac{S^{\ell}}{\langle 1\bar{2}\rangle ^{q_{\Phi}-\frac{1}{2}(\Delta+\ell)} (\langle 1\bar{0}\rangle \langle 0\bar{2}\rangle)^{\frac{1}{2}(\Delta+\ell)}}.\label{eq:Chiral3PF1}
\end{equation}
This is not manifestly of the form~(\ref{eq:AAO}).  However, we can render~(\ref{eq:Chiral3PF1}) in the form we desire by using the identity
\begin{equation}
\p{\frac{\<1\bar 2\>}{\langle 1\bar{0}\rangle\langle 0\bar{2}\rangle}}^{2\de}
=
\p{\frac{\langle 1\bar{2}\rangle\langle 2\bar{1}\rangle}{\langle 1\bar{0}\rangle \langle 0\bar{2}\rangle\langle 0\bar{1}\rangle \langle 2\bar{0}\rangle}}^\de(1-2\de z+2\de^2 z^{2}),
\label{eq:specialidentity}
\end{equation}
which follows readily from the definition~(\ref{eq:z}), together with the fact that $z$ is nilpotent of degree three.
Specializing to $\de=\frac{\De+\ell}{4}$, we obtain
\begin{align}
\langle \Phi(1,\bar{1})\Phi^{\dagger}(2,\bar{2})&\mathcal{O}(0,\bar{0},\cS,\bar{\cS})\rangle\nn\\
&=\lambda_{\Phi\Phi^{\dagger}\mathcal{O}}\frac{\left[1-\frac{1}{2}(\Delta+\ell)z+\frac{1}{8}(\Delta+\ell)(\Delta-\ell)z^{2}\right]S_{-}^{\ell}+\ell S_{+}S_{-}^{\ell-1}}{\langle 1\bar{2}\rangle ^{q_{\Phi}-\frac{1}{4}(\Delta+\ell)}\langle 2\bar{1}\rangle ^{-\frac{1}{4}(\Delta+\ell)}(\langle 0\bar{1}\rangle \langle 1\bar{0}\rangle \langle 0\bar{2}\rangle \langle 2\bar{0}\rangle)^{\frac{1}{4}(\Delta+\ell)}}.\label{eq:Chiral3PF2}
\end{align}
Eq.~(\ref{eq:Chiral3PF2}) is now explicitly consistent with Eq.~(\ref{eq:AAO}).  The relative ratio between the four coefficients $\lambda_{\Phi\Phi^{\dagger}\mathcal{O}}^{(i)}$ is fixed by the chirality condition $\bar{\mathcal{D}}_{\dot{a}}\Phi=0$. This alternative form will be useful when taking the chiral limit of our general superconformal block.

\subsubsection{Conserved Currents}

A conserved current superfield $\mathcal{J}$ has $q_{\mathcal{J}}=\bar{q}_{\mathcal{J}}=1$. This gives a three-point function
\begin{equation}
\langle \mathcal{J}_{1}(1,\bar{1})\mathcal{J}_{2}(2,\bar{2})\mathcal{O}(0,\bar{0},\cS,\bar{\cS})\rangle =\frac{\left(\lambda_{\mathcal{J}_{1}\mathcal{J}_{2}O}^{(0)}+\lambda_{\mathcal{J}_{1}\mathcal{J}_{2}O}^{(1)}z+\lambda_{\mathcal{J}_{1}\mathcal{J}_{2}O}^{(2)}z^{2}\right)S_{-}^{\ell}+\lambda_{\mathcal{J}_{1}\mathcal{J}_{2}O}^{(3)}S_{+}S_{-}^{\ell-1}}{(\langle 1\bar{2}\rangle \langle 2\bar{1}\rangle)^{1-\frac{1}{4}(\Delta+\ell)}(\langle 0\bar{1}\rangle \langle 1\bar{0}\rangle \langle 0\bar{2}\rangle \langle 2\bar{0}\rangle)^{\frac{1}{4}(\Delta+\ell)}}.\label{eq:JJO}
\end{equation}
These coefficients are related to each other by the current conservation conditions $\mathcal{D}^2 \mathcal{J}_{1,2}=\bar{\mathcal{D}}^2 \mathcal{J}_{1,2}=0$, which impose the constraints
\begin{eqnarray}
\lambda_{\mathcal{J}_{1}\mathcal{J}_{2}\cO}^{(2)} & = & \frac{1}{8}(4+\ell-\Delta)(\Delta+\ell) \lambda_{\mathcal{J}_{1}\mathcal{J}_{2}\cO}^{(0)},\nonumber \\
\lambda_{\mathcal{J}_{1}\mathcal{J}_{2}\cO}^{(3)} & = & -\frac{2(\Delta-2)}{\Delta+\ell} \lambda_{\mathcal{J}_{1}\mathcal{J}_{2}\cO}^{(1)}.\label{eq:Conservation}
\end{eqnarray}
However, the ratio between $\lambda_{\mathcal{J}_{1}\mathcal{J}_{2}O}^{(0)}$ and $\lambda_{\mathcal{J}_{1}\mathcal{J}_{2}O}^{(1)}$ remains arbitrary.

\subsection{Shadow Operator and Correlation Functions}

In order to calculate superconformal blocks we will also need the shadow three point function $\<\tl \cO \cB_1 \cB_2^\dag\>$, where $\cB_{1,2}$ have superconformal weight $(q_{\mathcal{B}},\bar{q}_{\mathcal{B}})$. This is given by applying Eq.~(\ref{Shadow}):
\begin{eqnarray}
\langle \tl{\mathcal{O}}(0,\bar 0,\cT,\bar{\cT})\mathcal{B}_{1}(3,\bar 3)\mathcal{B}_{2}^{\dag}(4,\bar 4)\rangle  & = & \int D[5,\bar 5]\frac{1}{\langle 0\bar{5}\rangle ^{1-\frac{1}{2}(\Delta-\ell)}\langle 5\bar{0}\rangle ^{1-\frac{1}{2}(\Delta-\ell)}}\times\nonumber \\
 &  & \langle \mathcal{O}(5,\bar{5}, 5\bar{\cT},\bar{5}\cT)\mathcal{B}_{1}(3,\bar 3)\mathcal{B}_{2}^\dag(4,\bar 4)\rangle .\label{eq:Shadow3PFDef}
\end{eqnarray}
By superconformal symmetry, it must take the form
\begin{equation}
\langle \tl{\mathcal{O}}(0,\bar 0,\cT,\bar{\cT})\mathcal{B}_{1}(3,\bar 3)\mathcal{B}_{2}^\dag(4,\bar 4)\rangle
 =\frac{\left(\lambda_{\cB_1\cB_2^\dag\tl{\cO}}^{(0)}+\lambda_{\cB_1\cB_2^\dag\tl{\cO}}^{(1)}\tl{z}+\lambda_{\cB_1\cB_2^\dag\tl{\cO}}^{(2)}\tl{z}^{2}\right)T_{-}^{\ell}+\lambda_{\cB_1\cB_2^\dag\tl{\cO}}^{(3)}T_{+}T_{-}^{\ell-1}}{\langle 3\bar{4}\rangle ^{q_{\cB}+\frac{1}{4}(\Delta-\ell-2)}\langle 4\bar{3}\rangle ^{\bar{q}_{\cB}+\frac{1}{4}(\Delta-\ell-2)}(\langle 0\bar{3}\rangle \langle 3\bar{0}\rangle \langle 0\bar{4}\rangle \langle 4\bar{0}\rangle )^{-\frac{1}{4}(\Delta-\ell-2)}},\label{eq:Shadow3PFForm}
\end{equation}
where $T_{\pm}$ is defined analogously to $S_{\pm}$, and $\tl{z} = z \big|_{1\rightarrow 3, 2 \rightarrow 4}$.

The shadow coefficients $\lambda_{\cB_1\cB_2^\dag\tl{\cO}}^{(i)}$ are linearly
related to the coefficients of the original operator $\lambda_{\cB_1\cB_2^\dag\cO}^{(i)}$.
The relation between the coefficients may be determined by explicitly computing the integral in Eq.~(\ref{eq:Shadow3PFDef}). The coefficients can also be uniquely fixed by requiring that the linear transformation respects the constraints for current and chiral 3-point functions and yields a partial wave consistent with unitarity. We give this argument explicitly in Appendix~\ref{app:shadowtransformation}.
The resulting linear transformation is given by
\begin{eqnarray}
\left( \begin{array}{c}
\lambda_{\cB_1\cB_2^\dag\tl{\cO}}^{(0)} \\
\lambda_{\cB_1\cB_2^\dag\tl{\cO}}^{(2)} 
\end{array} \right)
& = & 
\left( \begin{array}{cc}
\frac{1}{\Delta} & -\frac{8(\Delta-1)}{\Delta(\Delta+\ell)^{2}} \\
- \frac{(\Delta-1)(\Delta-\ell-2)^{2}}{8\Delta} & \frac{(\Delta-\ell-2)^{2}}{\Delta(\Delta+\ell)^{2}}
\end{array} \right)
\left( \begin{array}{c}
\lambda_{\cB_1\cB_2^\dag\cO}^{(0)} \\
\lambda_{\cB_1\cB_2^\dag\cO}^{(2)}
\end{array} \right) \nn
\\
\left( \begin{array}{c}
\lambda_{\cB_1\cB_2^\dag\tl{\cO}}^{(1)} \\
\lambda_{\cB_1\cB_2^\dag\tl{\cO}}^{(3)} 
\end{array} \right)
& = & 
 \left( \begin{array}{cc}
 \frac{(\Delta-\ell-2)^{2}}{(\Delta+\ell)^{2}} & 0 \\
\frac{4\ell(\Delta-1)}{(\Delta+\ell)^{2}} & 1
\end{array} \right)
\left( \begin{array}{c}
\lambda_{\cB_1\cB_2^\dag\cO}^{(1)} \\
\lambda_{\cB_1\cB_2^\dag\cO}^{(3)}
\end{array} \right)\label{eq:ShadowTransformation}
\end{eqnarray}
Here we have normalized the shadow transformation so that applying it twice takes the three-point function coefficients back to themselves.

\section{Partial Wave Computation}
\label{sec:PartialWave}

Now we will turn our attention to the four-point function,
\begin{equation}
\langle \mathcal{A}_{1}(1,\bar{1})\mathcal{A}_{2}^{\dagger}(2,\bar{2})\mathcal{B}_{1}(3,\bar{3})\mathcal{B}_{2}^{\dagger}(4,\bar{4})\rangle ,\label{4PF}
\end{equation}
\noindent where as in the previous section the superfields $\mathcal{A}_{1,2}$ and $\mathcal{B}_{1,2}$
are Lorentz scalars with superconformal weights $\mathcal{A}_{1,2}\sim(q_{\mathcal{A}},\bar{q}_{\mathcal{A}})$
and $\mathcal{B}_{1,2}\sim(q_{\mathcal{B}},\bar{q}_{\mathcal{B}})$.

We wish to compute the superconformal partial wave $\mathcal{W}_{\mathcal{O}}$
corresponding to the exchange of a real, spin-$\ell$ superfield $\mathcal{O}\sim(\frac{\ell}{2},\frac{\ell}{2},\frac{\De}{2},\frac{\De}{2})$
in the $(12)(34)$-channel.  Up to overall normalization, this is given by inserting the projector~(\ref{Projector})
into the four-point function,
\begin{eqnarray}
\mathcal{W}_{\mathcal{O}}&\propto&\langle \mathcal{A}_{1}\mathcal{A}_{2}^{\dagger}\left|\mathcal{O}\right|\mathcal{B}_{1}\mathcal{B}_{2}^{\dagger}\rangle \nonumber\\
&=&\int D[0,\bar{0}]\langle \mathcal{A}_{1}\mathcal{A}_{2}^{\dagger}\mathcal{O}_{\ell}(0,\bar{0},\cS,\bar{\cS})\rangle \overleftrightarrow{\mathcal{D}_{\ell}} \langle \tl{\mathcal{O}}_{\ell}(0,\bar{0},\cT,\bar{\cT})\mathcal{B}_{1}\mathcal{B}_{2}^{\dagger}\rangle \nonumber\\
&=&\frac{1}{\langle 1\bar{2}\rangle ^{q_{\mathcal{A}}-\frac14(\De+\ell)}\langle 2\bar{1}\rangle ^{\bar{q}_{\mathcal{A}}-\frac14(\De+\ell)}\langle 3\bar{4}\rangle ^{q_{\mathcal{B}}+\frac14(\De-\ell-2)}\langle 4\bar{3}\rangle ^{\bar{q}_{\mathcal{B}}+\frac14(\De-\ell-2)}} \times \nonumber\\
&& \hspace{10mm} \int D[0,\bar{0}]\frac{\mathcal{N}_{\ell}^{(\text{full})}}{(\langle 0\bar{1}\rangle \langle 1\bar{0}\rangle \langle 0\bar{2}\rangle \langle 2\bar{0}\rangle )^{\frac14(\De+\ell)}(\langle 0\bar{3}\rangle \langle 3\bar{0}\rangle \langle 0\bar{4}\rangle \langle 4\bar{0}\rangle )^{-\frac14(\De-\ell-2)}} \label{W1} 
\end{eqnarray}
\noindent where

\begin{eqnarray}
\mathcal{N}_{\ell}^{(\text{full})}&=&
\left[(\lambda^{(0)}_{\cA_1\cA_2^\dag\cO}+\lambda^{(1)}_{\cA_1\cA_2^\dag\cO}z+\lambda^{(2)}_{\cA_1\cA_2^\dag\cO}z^{2})S_{-}^{\ell}+\lambda^{(3)}_{\cA_1\cA_2^\dag\cO}S_{+}S_{-}^{\ell-1}\right] \nonumber\\
&& \overleftrightarrow{\mathcal{D}_{\ell}}
\left[(\lambda^{(0)}_{\cB_1\cB_2^\dag\tl{\cO}}+\lambda^{(1)}_{\cB_1\cB_2^\dag\tl{\cO}}\tl{z}+\lambda^{(2)}_{\cB_1\cB_2^\dag\tl{\cO}}\tl{z}^{2})T_{-}^{\ell}+\lambda^{(3)}_{\cB_1\cB_2^\dag\tl{\cO}}T_{+}T_{-}^{\ell-1}\right],\label{NFull1}
\end{eqnarray}

\noindent and we have introduced the shorthand notation
\begin{equation}
\overleftrightarrow{\mathcal{D}_{\ell}}\equiv\frac{1}{\ell!^{4}}(\partial_{\cS}0\partial_{\cT})^{\ell}(\partial_{\bar{\cS}}\bar{0}\partial_{\bar{\cT}})^{\ell}.\label{Deriv}
\end{equation}

We will not attempt to evaluate the integral~(\ref{W1}) in
full generality. Rather, as in \cite{Fitzpatrick:2014oza}, we will focus
on the case where the superfields in the four-point
function~(\ref{4PF}) are restricted to their lowest component
field. We refer to these superfields as the ``external'' fields,
in constrast to the ``exchanged'' operator $\mathcal{O}$. Setting
the external fields to their lowest component means setting their
Grassman coordinates, 
\begin{equation}
\theta_{\mathrm{ext}}\equiv\left\{ \theta_{i},\bar{\theta}_{i},\quad i=1,\dots4\right\} ,\label{Theta Ext}
\end{equation}
\noindent all to zero. This restriction is still of much interest,
because the exchanged operator $\mathcal{O}$ remains a full-fledged
superfield, and its associated partial wave is an essential ingredient
for supersymmetric bootstrap applications.  Note that setting $\theta_{\mathrm{ext}}$ to zero brings the integrand in~(\ref{W1}) to a form where Eq.~(\ref{eq:specialintegralformula}) applies.

After setting $\theta_{\mathrm{ext}}=0$, several terms in $\cN_{\ell}^{(\text{full})}$ vanish. In particular, $z$, $\tl{z}$, $S_{+}$, and
$T_{+}$ are all proportional to $\theta_{0}\bar{\theta}_{0}$, so%
\footnote{In our index-free formalism, $S_{+}$ and $T_{+}$ represent matrices.
Their proportionality to $\theta_{0}\bar{\theta}_{0}$ when $\theta_{\mathrm{ext}}=0$
can be seen, for instance, by going to a frame with $x_{1}\rightarrow0$,
$x_{2}\rightarrow\infty$ for $S_{+}$ and $x_{3}\rightarrow0$, $x_{4}\rightarrow\infty$
for $T_{+}$.%
}
\begin{equation}
\left.\left\{ z\tl{z}^{2},\quad z^{2}\tl{z},\quad z^{2}\tl{z}^{2},\quad z^{2}T_{+},\quad\tl{z}^{2}S_{+}\right\} \right|_{\theta_{\mathrm{ext}}=0}=0.\label{Simplify1}
\end{equation}
\noindent Therefore,
\begin{eqnarray}
\cN_{\ell}^{(\text{full})}&=&S_{-}^{\ell}\overleftrightarrow{\mathcal{D}_{\ell}}T_{-}^{\ell}\left[\lambda^{(0)}_{\cA_1\cA_2^\dag\cO}\lambda^{(0)}_{\cB_1\cB_2^\dag\tl{\cO}}+\lambda^{(1)}_{\cA_1\cA_2^\dag\cO}\lambda^{(0)}_{\cB_1\cB_2^\dag\tl{\cO}}z+\lambda^{(0)}_{\cA_1\cA_2^\dag\cO}\lambda^{(1)}_{\cB_1\cB_2^\dag\tl{\cO}}\tl{z} 
 \right. \nonumber \\
&&\hspace{17mm} \left. +\lambda^{(2)}_{\cA_1\cA_2^\dag\cO}\lambda^{(0)}_{\cB_1\cB_2^\dag\tl{\cO}}z^{2}+\lambda^{(0)}_{\cA_1\cA_2^\dag\cO}\lambda^{(2)}_{\cB_1\cB_2^\dag\tl{\cO}}\tl{z}^{2}+\lambda^{(1)}_{\cA_1\cA_2^\dag\cO}\lambda^{(1)}_{\cB_1\cB_2^\dag\tl{\cO}}z\tl{z}\right] \nonumber \\
&+&S_{-}^{\ell}\overleftrightarrow{\mathcal{D}_{\ell}}T_{+}T_{-}^{\ell-1}\left[\lambda^{(0)}_{\cA_1\cA_2^\dag\cO}\lambda^{(3)}_{\cB_1\cB_2^\dag\tl{\cO}}+\lambda^{(1)}_{\cA_1\cA_2^\dag\cO}\lambda^{(3)}_{\cB_1\cB_2^\dag\tl{\cO}}z\right] \nonumber \\
&+&S_{+}S_{-}^{\ell-1}\overleftrightarrow{\mathcal{D}_{\ell}}T_{-}^{\ell}\left[\lambda^{(3)}_{\cA_1\cA_2^\dag\cO}\lambda^{(0)}_{\cB_1\cB_2^\dag\tl{\cO}}+\lambda^{(3)}_{\cA_1\cA_2^\dag\cO}\lambda^{(1)}_{\cB_1\cB_2^\dag\tl{\cO}}\tl{z}\right] \nonumber \\
&+&S_{+}S_{-}^{\ell-1}\overleftrightarrow{\mathcal{D}_{\ell}}T_{+}T_{-}^{\ell-1}\left[\lambda^{(3)}_{\cA_1\cA_2^\dag\cO}\lambda^{(3)}_{\cB_1\cB_2^\dag\tl{\cO}}\right]+\dots\label{NFull2}
\end{eqnarray}
\noindent where the dots denote terms that vanish when $\theta_{\mathrm{ext}}=0$.

The supertraces in Eq.~(\ref{W1})
reduce as
\begin{equation}
\left.\langle \bar{i}j\rangle \right|_{\theta_{\mathrm{ext}}=0}=-X_{i}\cdot X_{j}\equiv\frac{1}{2}X_{ij},
\end{equation}
\noindent where the $X_{i}\in\mathbb{C}^{6}$ on the right are non-supersymmetric
embedding vectors. With these observations in mind, the prescription
in Eq.~(\ref{eq:specialintegralformula}) then gives us
\begin{equation}
\mathcal{\left.W_{O}\right|}_{\theta_{\mathrm{ext}}=0}\propto\frac{1}{X_{12}^{\Delta_{\mathcal{A}}-\frac12(\De+\ell)}X_{34}^{\Delta_{\mathcal{B}}+\frac12(\De-\ell-2)}}\int D^{4}X_{0}\left.\partial_{\bar{0}}^{2}\frac{\cN_{\ell}^{(\text{full})}}{D_{\ell}}\right|_{\bar{0}=0},
\label{W2}
\end{equation}
\noindent where 
\begin{equation}
\frac{1}{D_{\ell}}\equiv\frac{1}{(X_{10}X_{1\bar{0}}X_{20}X_{2\bar{0}})^{\frac14(\De+\ell)}(X_{30}X_{3\bar{0}}X_{40}X_{4\bar{0}})^{-\frac14(\De-\ell-2)}}.\label{DL}
\end{equation}

To compute the derivative in Eq.~(\ref{W2}), it is natural to introduce
the object
\begin{equation}
\cN_{\ell}\equiv(\bar{\cS}1\bar{2}\cS)^{\ell}\overleftrightarrow{\mathcal{D}_{\ell}}(\bar{\cT}3\bar{4}\cT)^{\ell}\label{NL}
\end{equation}
\noindent This is the same quantity we encountered in our computation
of superconformal blocks for the chiral four-point function in~\cite{Fitzpatrick:2014oza}.
In Appendix~\ref{app:Derivs}, we list several properties of $\cN_{\ell}$. For now, we simply note that when $\theta_{\mathrm{ext}}=0$, 
$\cN_{\ell}(\cX_0,\bar{\cX}_0)\rightarrow \cN_{\ell}(X_0,X_{\bar{0}})$, and when $\bar{0}=0$
\begin{equation}
\left.\cN_{\ell}(X_0,X_{\bar{0}})\right|_{\bar{0}=0}= a_{\ell} (X_{12}X_{34}X_{10}X_{20}X_{30}X_{40})^{\frac{\ell}{2}}(-1)^\ell C_{\ell}^{(1)}(t_{0})\label{NL0},
\end{equation}
\noindent where $C_{\ell}^{(\lambda)}(x)$ are the Gegenbauer polynomials, $a_{\ell} = 2^{-6\ell}$,
and 
\begin{equation}
t_{0}=-\frac{X_{13}X_{20}X_{40}}{2\sqrt{X_{10}X_{20}X_{30}X_{40}X_{12}X_{34}}}-(1\leftrightarrow2)-(3\leftrightarrow4).
\label{t0}
\end{equation}
\noindent In our notation, the $(3\leftrightarrow4)$ acts on both the first term and the $(1\leftrightarrow2)$ term. 

It is natural to introduce $\cN_{\ell}$, because $\cN_{\ell}^{(\text{full})}$
can be written in terms of $\cN_{\ell}$ up to coordinate exchanges
that take $(1\leftrightarrow2)$ and/or $(3\leftrightarrow4)$.
In particular, recalling Eq.~(\ref{eq:3PFBasisChange}) for $S_{-}^{\ell}$ and
$S_{+}S_{-}^{\ell-1}$, we have
\begin{eqnarray}
S_{-}^{\ell}\overleftrightarrow{\mathcal{D}_{\ell}}T_{-}^{\ell}&=&\frac{\cN_{\ell}}{4\langle 1\bar{2}\rangle ^{\ell}\langle 3\bar{4}\rangle ^{\ell}}+(-1)^{\ell}(1\leftrightarrow2)+(-1)^{\ell}(3\leftrightarrow4) \nonumber\\
S_{-}^{\ell}\overleftrightarrow{\mathcal{D}_{\ell}}T_{+}T_{-}^{\ell-1}&=&\frac{\cN_{\ell}}{4\ell\langle 1\bar{2}\rangle ^{\ell}\langle 3\bar{4}\rangle ^{\ell}}+(-1)^{\ell}(1\leftrightarrow2)-(-1)^{\ell}(3\leftrightarrow4) \nonumber\\
S_{+}S_{-}^{\ell-1}\overleftrightarrow{\mathcal{D}_{\ell}}T_{-}^{\ell}&=&\frac{\cN_{\ell}}{4\ell\langle 1\bar{2}\rangle ^{\ell}\langle 3\bar{4}\rangle ^{\ell}}-(-1)^{\ell}(1\leftrightarrow2)+(-1)^{\ell}(3\leftrightarrow4) \nonumber\\
S_{+}S_{-}^{\ell-1}\overleftrightarrow{\mathcal{D}_{\ell}}T_{+}T_{-}^{\ell-1}&=&\frac{\cN_{\ell}}{4\ell^{2}\langle 1\bar{2}\rangle ^{\ell}\langle 3\bar{4}\rangle ^{\ell}}-(-1)^{\ell}(1\leftrightarrow2)-(-1)^{\ell}(3\leftrightarrow4). 
\label{Symm}
\end{eqnarray}%
\noindent Similar formulas with the left-hand side multiplied by $z$'s
and $\tl{z}$'s are easily obtained by remembering that $z$ is
antisymmetric in $(1\leftrightarrow2)$ and independent
of $3$ and $4$, and vice versa for $\tl{z}$. 

At this point, an important simplifying observation is that $\mathcal{\left.W_{O}\right|}_{\theta_{\mathrm{ext}}=0}$
is invariant under the simultaneous coordinate interchange $1\leftrightarrow2$,
$3\leftrightarrow4$. Therefore, in Eq.~(\ref{W2}), any piece of $\cN_{\ell}^{(\text{full})}|_{\theta_{\mathrm{ext}}=0}$ 
that is antisymmetric under this interchange must have vanishing contribution.
By Eqs.~(\ref{Symm}), this is true for the following terms,
\[
zS_{-}^{\ell}\overleftrightarrow{\mathcal{D}_{\ell}}T_{-}^{\ell},\qquad\tl{z}S_{-}^{\ell}\overleftrightarrow{\mathcal{D}_{\ell}}T_{-}^{\ell},\qquad S_{-}^{\ell}\overleftrightarrow{\mathcal{D}_{\ell}}T_{+}T_{-}^{\ell-1},\qquad S_{+}S_{-}^{\ell-1}\overleftrightarrow{\mathcal{D}_{\ell}}T_{-}^{\ell},
\]
\noindent and so we can ignore these terms from the outset.%
\footnote{We have checked explicitly that each of these terms has vanishing
contribution in Eq.~(\ref{W2}).%
} This leaves us with
\begin{eqnarray}
\cN_{\ell}^{(\text{full})}&=&S_{-}^{\ell}\overleftrightarrow{\mathcal{D}_{\ell}}T_{-}^{\ell}\left[\lambda^{(0)}_{\cA_1\cA_2^\dag\cO}\lambda^{(0)}_{\cB_1\cB_2^\dag\tl{\cO}}+\lambda^{(2)}_{\cA_1\cA_2^\dag\cO}\lambda^{(0)}_{\cB_1\cB_2^\dag\tl{\cO}}z^{2}
 \right. \nonumber \\
&&\hspace{17mm} \left. +\lambda^{(0)}_{\cA_1\cA_2^\dag\cO}\lambda^{(2)}_{\cB_1\cB_2^\dag\tl{\cO}}\tl{z}^{2}+\lambda^{(1)}_{\cA_1\cA_2^\dag\cO}\lambda^{(1)}_{\cB_1\cB_2^\dag\tl{\cO}}z\tl{z} \right] \nonumber \\
&+&S_{-}^{\ell}\overleftrightarrow{\mathcal{D}_{\ell}}T_{+}T_{-}^{\ell-1}\left[\lambda^{(1)}_{\cA_1\cA_2^\dag\cO}\lambda^{(3)}_{\cB_1\cB_2^\dag\tl{\cO}}z\right] \nonumber \\
&+&S_{+}S_{-}^{\ell-1}\overleftrightarrow{\mathcal{D}_{\ell}}T_{-}^{\ell}\left[\lambda^{(3)}_{\cA_1\cA_2^\dag\cO}\lambda^{(1)}_{\cB_1\cB_2^\dag\tl{\cO}}\tl{z}\right] \nonumber \\
&+&S_{+}S_{-}^{\ell-1}\overleftrightarrow{\mathcal{D}_{\ell}}T_{+}T_{-}^{\ell-1}\left[\lambda^{(3)}_{\cA_1\cA_2^\dag\cO}\lambda^{(3)}_{\cB_1\cB_2^\dag\tl{\cO}}\right]+\dots\label{NFull3}
\end{eqnarray}%
\noindent where the dots denote terms that do not contribute to $\mathcal{\left.W_{O}\right|}_{\theta_{\mathrm{ext}}=0}$.

The remainder of the computation is straightforward. One inserts Eq.~(\ref{NFull3})
into Eq.~(\ref{W2}) and computes the $\partial_{\bar{0}}^{2}$ derivatives.
This results in a sum over various conformal
integrals, which are evaluated using the result of~\cite{DO1,SimmonsDuffin:2012uy}:
\begin{eqnarray}
\left.\int D^{4}X_{0}\frac{(-1)^{\ell} C_{\ell}^{(1)}(t_{0})}{X_{10}^{\frac{\Delta+\Delta_{12}}{2}}X_{20}^{\frac{\Delta-\Delta_{12}}{2}}X_{30}^{\frac{\tl{\Delta}+\Delta_{34}}{2}}X_{40}^{\frac{\tl{\Delta}-\Delta_{34}}{2}}}\right|_{M=1} && \nonumber \\
&&\hspace{-55mm} =\xi_{\Delta,\tl{\Delta},\Delta_{34},\ell}\left(\frac{X_{14}}{X_{13}}\right)^{\frac{\Delta_{34}}{2}}\left(\frac{X_{24}}{X_{14}}\right)^{\frac{\Delta_{12}}{2}}X_{12}^{-\frac{\Delta}{2}}X_{34}^{-\frac{\tl{\Delta}}{2}}g_{\Delta,\ell}^{\Delta_{12},\Delta_{34}}(u,v)
\label{ConfInt}
\end{eqnarray}
\noindent where
\begin{equation}
\xi_{\Delta,\tl{\Delta},\Delta_{34},\ell}\equiv\frac{\pi^{2}\Gamma(\tl{\Delta}+\ell-1)\Gamma(\frac{\Delta-\Delta_{34}+\ell}{2})\Gamma(\frac{\Delta+\Delta_{34}+\ell}{2})}{(2-\Delta)\Gamma(\Delta+\ell)\Gamma(\frac{\tl{\Delta}-\Delta_{34}+\ell}{2})\Gamma(\frac{\tl{\Delta}+\Delta_{34}+\ell}{2})},\label{xi}
\end{equation}
\noindent and $g_{\Delta,\ell}^{\Delta_{12},\Delta_{34}}(u,v)$ are
the usual non-supersymmetric conformal blocks given by\footnote{Our definition of $g_{\Delta,\ell}^{\Delta_{12},\Delta_{34}}(u,v)$ differs by factors of $(-2)^\ell$ and $(-1)^\ell$ from the normalizations used in \cite{Poland:2010wg,DO1} and \cite{SimmonsDuffin:2012uy}, respectively.}
\begin{eqnarray}
g_{\Delta,\ell}^{\Delta_{12},\Delta_{34}}(u,v)&=&\frac{z\bar{z}}{z-\bar{z}}\left[k_{\Delta+\ell}(z)k_{\Delta-\ell-2}(\bar{z})-(z\leftrightarrow\bar{z})\right]\label{g}, \\
k_{\beta}(x)&=&x^{\frac{\beta}{2}}{}_{2}F_{1}\left(\frac{\beta-\Delta_{12}}{2},\frac{\beta+\Delta_{34}}{2},\beta,x\right),\nn \\
u&=&z \bar{z}, \hspace{10mm} v=(1-z)(1-\bar{z}).\nn
\end{eqnarray}
Here, $\De_{ij}\equiv \De_i-\De_j$, and $g^{\De_{12},\De_{34}}_{\De,\ell}$ is the conformal block for exchange of an operator with dimension $\De$ and spin $\ell$ in a four point function of scalars with dimension $\De_i$.
The resulting expression for $\mathcal{\left.W_{O}\right|}_{\theta_{\mathrm{ext}}=0}$ is a linear combination of the $g_{\Delta,\ell}^{\Delta_{12},\Delta_{34}}$
(as expected). The corresponding superconformal block $\cG^{\mathcal{N}=1 | A_1 A_2^{\dagger} ; B_1 B_2^{\dagger}}_{\De,\ell}$ is simply related by
\begin{equation}
\cG^{\mathcal{N}=1 | A_1 A_2^{\dagger} ; B_1 B_2^{\dagger}}_{\De,\ell}=(X_{12})^{\Delta_{\mathcal{A}}}(X_{34})^{\Delta_{\mathcal{B}}}\mathcal{\left.W_{O}\right|}_{\theta_{\mathrm{ext}}=0}.
\end{equation}
\noindent Additional details of the calculation are given in Appendix~\ref{app:Derivs}.

\section{Results}
\label{sec:results}

After relating the shadow coefficients to the coefficients of the original operator using Eqs.~(\ref{eq:ShadowTransformation}), the computations described in the previous section give the result
\begin{eqnarray}
\mathcal{G}_{\De, \ell}^{\mathcal{N}=1 | A_1 A_2^{\dagger} ; B_1 B_2^{\dagger}} &=& \nonumber\\
 &  & \hspace{-20mm} \lambda_{\cA_1\cA_2^\dag\cO}^{(0)}\lambda_{\cB_1\cB_2^\dag\cO}^{(0)}g_{\Delta,\ell}+\frac{\lambda_{\cA_1\cA_2^\dag\cO}^{(1)}\lambda_{\cB_1\cB_2^\dag\cO}^{(1)}}{(\Delta+\ell)(\Delta+\ell+1)}g_{\Delta+1,\ell+1}\nonumber \\
 &  & \hspace{-20mm} +\frac{\left[\lambda_{\cA_1\cA_2^\dag\cO}^{(1)}+\frac{\ell+1}{\ell}\lambda_{\cA_1\cA_2^\dag\cO}^{(3)}\right]\left[\lambda_{\cB_1\cB_2^\dag\cO}^{(1)}+\frac{\ell+1}{\ell}\lambda_{\cB_1\cB_2^\dag\cO}^{(3)}\right]}{(\Delta-\ell-1)(\Delta-\ell-2)}g_{\Delta+1,\ell-1}\nonumber \\
 &  & \hspace{-20mm} +\frac{\left[(\Delta+\ell)^{2}\lambda_{\cA_1\cA_2^\dag\cO}^{(0)}-8(\Delta-1)\lambda_{\cA_1\cA_2^\dag\cO}^{(2)}\right]\left[(\Delta+\ell)^{2}\lambda_{\cB_1\cB_2^\dag\cO}^{(0)}-8(\Delta-1)\lambda_{\cB_1\cB_2^\dag\cO}^{(2)}\right]}{16\Delta^{2}(\Delta-\ell-1)(\Delta-\ell-2)(\Delta+\ell)(\Delta+\ell+1)}g_{\Delta+2,\ell},\nonumber\\\label{eq:BlockRes1}
\end{eqnarray}
where we have retained the overall OPE coefficient dependence to make it clear which structures contribute to each term.  Here, $g_{\De,\ell}=g^{0,0}_{\De,\ell}$ is the conformal block for external scalars with $\De_1=\De_2$ and $\De_3=\De_4$.  The different terms present above reflect the decomposition of the superconformal multiplet of $\cO$ into conformal multiplets \cite{Poland:2010wg}.

If we take $\cB_{2,1} = \cA_{1,2}$, then $\lambda_{\cB_1\cB_2^\dag\cO}^{(i)}=\lambda_{\cA_2\cA_1^\dag\cO}^{(i)}=\left(\lambda_{\cA_1\cA_2^\dag\cO}^{(i)}\right)^\dag$ and we obtain the superconformal block
\begin{eqnarray}
\mathcal{G}_{\De,\ell}^{\mathcal{N}=1 | A_1 A_2^{\dagger} ; A_2 A_1^{\dagger}} & = & \left|{\lambda_{\cA_1\cA_2^\dag\cO}^{(0)}}\right|^{2}g_{\Delta,\ell}+\frac{\left|{\lambda_{\cA_1\cA_2^\dag\cO}^{(1)}}\right|^{2}}{(\Delta+\ell)(\Delta+\ell+1)}g_{\Delta+1,\ell+1}\nonumber \\
 &  & +\frac{\left|\lambda_{\cA_1\cA_2^\dag\cO}^{(1)}+\frac{\ell+1}{\ell}\lambda_{\cA_1\cA_2^\dag\cO}^{(3)}\right|^{2}}{(\Delta-\ell-1)(\Delta-\ell-2)}g_{\Delta+1,\ell-1}\nonumber \\
 &  & +\frac{\left|(\Delta+\ell)^{2}\lambda_{\cA_1\cA_2^\dag\cO}^{(0)}-8(\Delta-1)\lambda_{\cA_1\cA_2^\dag\cO}^{(2)}\right|^{2}}{16 \Delta^{2}(\Delta-\ell-1)(\Delta-\ell-2)(\Delta+\ell)(\Delta+\ell+1)}g_{\Delta+2,\ell}.\label{eq:BlockRes2}
\end{eqnarray}
If there are no further constraints on $\lambda^{(i)}$, then $\mathcal{N}=1$
superconformal symmetry cannot fix the relative coefficients between
the supermultiplet of conformal blocks. However, additional symmetries or shortening 
conditions may impose interesting constraints on $\lambda^{(i)}$. For the four-point
function $\langle \phi\phi^{\dagger}\phi\phi^{\dagger}\rangle $,
where $\phi$ is the lowest component of a chiral multiplet $\Phi$, we
may plug in the three-point function coefficients in Eq.~(\ref{eq:Chiral3PF2})
to obtain: 
\begin{eqnarray}
\mathcal{G}_{\De,\ell}^{\mathcal{N}=1 | \phi \phi^\dag ; \phi \phi^\dag} & = & \left|\l_{\Phi \Phi^{\dagger} \cO}\right|^2 \left[ g_{\Delta,\ell} +\frac{(\Delta-\ell-2)}{4(\Delta-\ell-1)}g_{\Delta+1,\ell-1}+\frac{(\Delta+\ell)}{4(\Delta+\ell+1)}g_{\Delta+1,\ell+1} \right.\nonumber \\
 &  & \left. \qquad\qquad + \frac{(\Delta+\ell)(\Delta-\ell-2)}{16(\Delta+\ell+1)(\Delta-\ell-1)}g_{\Delta+2,\ell} \right] .\label{eq:BlockResChiral}
\end{eqnarray}
This agrees exactly with the previous results derived in~\cite{Poland:2010wg} and provides a nontrivial check for our formalism. 

Next let us consider the four-point function $\langle J_{1}J_{2}J_{3}J_{4}\rangle $,
where $J_{i}$ is the lowest component of a global symmetry current
multiplet $\cJ_{i}$. This case was considered recently in~\cite{Fortin:2011nq,Berkooz:2014yda}.
The conservation condition $\mathcal{D}^{2}\mathcal{J}_{i}=0$
imposes constraints on the three-point function coefficients as in Eqs.~(\ref{eq:Conservation}).
Plugging in these relations we get: 
\begin{eqnarray}
\mathcal{G}_{\De,\ell}^{\mathcal{N}=1 | J_{1}J_{2} ; J_{3}J_{4}} & = & \nonumber \\
 &  & \hspace{-20mm} {\lambda_{\cJ_{1}\cJ_{2}\cO}^{(0)}}{\lambda_{\cJ_{3}\cJ_{4}\cO}^{(0)}}\left[g_{\Delta,\ell}+\frac{1}{16}\frac{(\Delta-2)^{2}(\Delta+\ell)(\Delta-\ell-2)}{\Delta^{2}(\Delta+\ell+1)(\Delta-\ell-1)}g_{\Delta+2,\ell}\right]\nonumber \\
 &  & \hspace{-20mm} \!\!\!\!\!+{\lambda_{\cJ_{1}\cJ_{2}\cO}^{(1)}}{\lambda_{\cJ_{3}\cJ_{4}\cO}^{(1)}}\left[\frac{1}{(\Delta+\ell)(\Delta+\ell+1)}g_{\Delta+1,\ell+1}+\frac{(\ell+2)^{2}(\Delta-\ell-2)}{\ell^{2}(\Delta+\ell)^{2}(\Delta-\ell-1)}g_{\Delta+1,\ell-1}\right],\nonumber\\\label{eq:BlockResCurrent}
\end{eqnarray}
where the ratio $\frac{\lambda_{\cJ_{i}\cJ_{j}\cO}^{(1)}}{\lambda_{\cJ_{i}\cJ_{j}\cO}^{(0)}}$
is in general not fixed. 

When $J_1,\dots, J_4$ are identical conserved currents, then
the four-point function is symmetric under permutations $(x_{1}\leftrightarrow x_{2})$
or $(x_{3}\leftrightarrow x_{4})$. This further constrains
the three-point function coefficient. In particular, for even spin, $\lambda_{\cJ\cJ\cO}^{(1)}=\lambda_{\cJ\cJ\cO}^{(3)}=0$
and we have: 
\begin{equation}
\mathcal{G}_{\De,\ell, \text{even}}^{\mathcal{N}=1 | JJ ; JJ}=\left({\lambda_{\cJ\cJ\cO}^{(0)}}\right)^{2} \left[ g_{\Delta,\ell}+\frac{1}{16}\frac{(\Delta-2)^{2}(\Delta+\ell)(\Delta-\ell-2)}{\Delta^{2}(\Delta+\ell+1)(\Delta-\ell-1)}g_{\Delta+2,\ell} \right].\label{eq:BlockResJEven}
\end{equation}
For odd spin, $\lambda_{\cJ\cJ\cO}^{(0)}=\lambda_{\cJ\cJ\cO}^{(2)}=0$ and we have: 
\begin{equation}
\mathcal{G}_{\De,\ell,\text{odd}}^{\mathcal{N}=1 | JJ ; JJ}=\left({\lambda_{\cJ\cJ\cO}^{(1)}}\right)^{2} \left[\frac{1}{(\Delta+\ell)(\Delta+\ell+1)}g_{\Delta+1,\ell+1}+\frac{(\ell+2)^{2}(\Delta-\ell-2)}{\ell^{2}(\Delta+\ell)^{2}(\Delta-\ell-1)}g_{\Delta+1,\ell-1}\right].\label{eq:BlockResJOdd}
\end{equation}

For the four-point function $\langle J_{1}J_{2}\phi\phi^{\dagger}\rangle $,
we may plug in the conservation constraints for $\lambda_{\cA_1\cA_2^\dag\cO}^{(i)}$
and the chirality constraints for $\lambda_{\cB_1\cB_2^\dag\cO}^{(i)}$ and find:
\begin{eqnarray}
\mathcal{G}_{\De,\ell}^{\mathcal{N}=1 | J_1 J_2 ; \phi \phi^{\dagger}} & = & \lambda_{\cJ_{1}\cJ_{2} \cO}^{(0)} \lambda_{\Phi \Phi^{\dagger} \cO} \left[g_{\Delta,\ell}-\frac{1}{16}\frac{(\Delta-2)(\Delta+\ell)(\Delta-\ell-2)}{\Delta(\Delta+\ell+1)(\Delta-\ell-1)}g_{\Delta+2,\ell}\right]\nonumber \\
 &  & \hspace{-3mm} 
+\lambda_{\cJ_{1}\cJ_{2}\cO}^{(1)}\lambda_{\Phi \Phi^{\dagger} \cO}\left[-\frac{1}{2(\Delta+\ell+1)}g_{\Delta+1,\ell+1}+\frac{(\ell+2)(\Delta-\ell-2)}{2\ell(\Delta+\ell)(\Delta-\ell-1)}g_{\Delta+1,\ell-1}\right]. \nonumber\\\label{eq:BlockResCurrentChiral}
\end{eqnarray}
When $J_{1}$ and $J_{2}$ are identical currents, the even or odd
spin blocks will pick up different parts of this result after setting $\lambda_{\cJ\cJ\cO}^{(1)}=0$ or $\lambda_{\cJ\cJ\cO}^{(0)}=0$, respectively:
\begin{eqnarray}
\mathcal{G}_{\De,\ell,\text{even}}^{\mathcal{N}=1 | J J ; \phi \phi^{\dagger}} & = & \lambda_{\cJ\cJ\cO}^{(0)} \lambda_{\Phi \Phi^{\dagger} \cO} \left[g_{\Delta,\ell}-\frac{1}{16}\frac{(\Delta-2)(\Delta+\ell)(\Delta-\ell-2)}{\Delta(\Delta+\ell+1)(\Delta-\ell-1)}g_{\Delta+2,\ell}\right] \label{eq:BlockResCurrentChiralEven}
\end{eqnarray}
and
\begin{eqnarray}
\mathcal{G}_{\De,\ell,\text{odd}}^{\mathcal{N}=1 | J J ; \phi \phi^{\dagger}} & =  & \lambda_{\cJ\cJ\cO}^{(1)}\lambda_{\Phi \Phi^{\dagger} \cO}\left[-\frac{1}{2(\Delta+\ell+1)}g_{\Delta+1,\ell+1}+\frac{(\ell+2)(\Delta-\ell-2)}{2\ell(\Delta+\ell)(\Delta-\ell-1)}g_{\Delta+1,\ell-1}\right]. \nonumber\\\label{eq:BlockResCurrentChiralOdd}
\end{eqnarray}
The superconformal blocks in Eqs.~(\ref{eq:BlockResCurrent}-\ref{eq:BlockResCurrentChiralOdd}) are in agreement with the expressions in the most recent version of~\cite{Berkooz:2014yda} and the version of~\cite{Fortin:2011nq} to appear soon.

\section{Decomposition of $\cN=2$ Blocks into $\cN=1$ Blocks}
\label{sec:decomp}

We can get a nontrivial consistency check on the superconformal blocks derived in the previous section from decomposing $\cN=2$ superconformal blocks into $\cN=1$ superconformal blocks. Following closely the discussions of~\cite{Poland:2010wg,Fortin:2011nq}, we can consider $\cN=2$ global symmetry current multiplets $\varphi^{ij}$, which are $SU(2)_R$ triplets (neutral under $U(1)_R^{\cN=2}$) with dimension $\Delta=2$. The components $(\varphi^{11},\varphi^{(12)},\varphi^{22}) = (\phi, J,  \phi^{\dagger})$ are $\cN=1$ chiral, current, and anti-chiral, respectively. Then four-point functions $\<\varphi \varphi \varphi \varphi\>$ can be decomposed into contributions from different $\SU(2)_R$ channels, following the theory of Clebsch-Gordan coefficients, as
\be
\label{eq:decomp}
\mathcal{G}^{\cN=2 | \phi \phi^{\dagger} ; \phi \phi^{\dagger}} &=& A_0 + \frac12 A_1 + \frac16 A_2, \nn \\
\mathcal{G}^{\cN=2 | JJ ; JJ} &=& A_0 + \frac23 A_2, \nn\\
\mathcal{G}^{\cN=2 | JJ ; \phi \phi^{\dagger}} &=& A_0 - \frac13 A_2 .
\ee
Here the functions $A_R$ for $R=0,1,2$ reflect the contributions from each representation of $\SU(2)_R$ appearing in $\mathbf{3} \otimes \mathbf{3} = \mathbf{1} \oplus \mathbf{3} \oplus \mathbf{5}$. 

For the exchange of long multiplets $\cO$ of dimension $\Delta$, even spin $\ell$, and vanishing $R$ charge, Dolan and Osborn computed the contributions to be~\cite{Dolan:2001tt}
\be\label{eq:As}
A_0 &=& g_{\De,\ell}  + \frac{(\De+\ell+2)^2}{16(\De+\ell+1)(\De+\ell+3)} g_{\De+2,\ell+2}  + \frac{(\De-\ell)^2}{16(\De-\ell-1)(\De-\ell+1)} g_{\De+2,\ell-2}  \nn\\
&& + \frac{1}{12} g_{\De+2,\ell} + \frac{{(\De+\ell+2)^2(\De-\ell)^2}}{256(\De+\ell+1)(\De+\ell+3)(\De-\ell-1)(\De-\ell+1)}g_{\De+4,\ell}, \nn\\
A_1 &=& g_{\De+1,\ell+1}+g_{\De+1,\ell-1}+\frac{(\De+\ell+2)^2}{16(\De+\ell+1)(\De+\ell+3)}g_{\De+3,\ell+1} \nn\\
&&+ \frac{(\De-\ell)^2}{16(\De-\ell-1)(\De-\ell+1)}g_{\De+3,\ell-1}, \nn\\
A_2 &=& g_{\De+2,\ell}.
\ee

The chiral block decomposition using these results was performed in~\cite{Poland:2010wg}, with the result
\be
\mathcal{G}^{\cN=2 | \phi \phi^{\dagger} ; \phi \phi^{\dagger}}_{\De,\ell} &=& \mathcal{G}^{\cN=1 | \phi \phi^{\dagger} ; \phi \phi^{\dagger}}_{\De,\ell} + \frac{(\De-\ell)}{4(\De-\ell-1)} \mathcal{G}^{\cN=1 | \phi \phi^{\dagger} ; \phi \phi^{\dagger}}_{\De+1,\ell-1} + \frac{(\De+\ell+2)}{4(\De+\ell+1)}  \mathcal{G}^{\cN=1 | \phi \phi^{\dagger} ; \phi \phi^{\dagger}}_{\De+1,\ell+1} \nn \\
&& + \frac{(\De-\ell)(\De+\ell+2)}{16(\De-\ell-1)(\De+\ell+1)} \mathcal{G}^{\cN=1 | \phi \phi^{\dagger} ; \phi \phi^{\dagger}}_{\De+2,\ell} .
\ee

Now we extend this result to the decomposition of $\mathcal{G}^{\cN=2 | JJ ; JJ}$ and $\mathcal{G}^{\cN=2 | JJ ; \phi \phi^{\dagger}}$. As explained in~\cite{Berkooz:2014yda}, the $J \times J$ OPE can in general contain the descendants of unprotected $\cN=1$ primaries with vanishing $R$-charge and $(j,\bar{j}) = (\frac{\ell\pm1}{2},\frac{\ell\mp 1}{2})$. In the present context, such operators arise as $\cN=2$ descendants of the schematic form $Q_2 \bar{Q}_2 \cO$, with dimension $\Delta+1$. Because these operators have only one $\cN=1$ descendant (which is a conformal primary) with vanishing $R$-charge and integer spin, up to a normalization factor the corresponding superconformal block is just a conformal block
\be\label{eq:newblock}
\mathcal{G}^{\cN=1 | JJ ; JJ}_{\De+1,(\frac{\ell \pm 1}{2},\frac{\ell \mp 1}{2} )} &\propto& g_{\De+2,\ell} . 
\ee
Taking these contributions into account, we expect a decomposition of the form
\be
\mathcal{G}^{\cN=2 | JJ ; JJ}_{\De,\ell,\text{even}} &=& \mathcal{G}^{\cN=1 | JJ ; JJ}_{\De,\ell,\text{even}} + N(\De,\ell) \mathcal{G}^{\cN=1 | JJ ; JJ}_{\De+1,\ell - 1,\text{odd}} + J(\De,\ell) \mathcal{G}^{\cN=1 | JJ ; JJ}_{\De+1,\ell + 1,\text{odd}} \nn \\
&& + D(\De,\ell) \mathcal{G}^{\cN=1 | JJ ; JJ}_{\De+2,\ell,\text{even}} + B(\De,\ell) g_{\De+2,\ell} ,
\ee
where we have absorbed the overall $\lambda^2$ factors in Eqs.~(\ref{eq:BlockResJEven}) and (\ref{eq:BlockResJOdd}) into the functions $N,J,D$. Matching to Eqs.~(\ref{eq:decomp}) and (\ref{eq:As}) gives
\be
N(\De,\ell) &=& \frac{(\ell-1)^2(\De-\ell)(\De+\ell)^2}{16(\ell+1)^2(\De-\ell-1)},\\
J(\De,\ell) &=& \frac{(\De+\ell+2)^3}{16(\De+\ell+1)},\\
D(\De,\ell) &=&\frac{(\De+2)^2(\De-\ell)(\De+\ell+2)}{16\De^2(\De-\ell-1)(\De+\ell+1)},\\
B(\De,\ell) &=&\frac{\ell (\ell+2) (\De+1)(\De-1)}{2 (\ell+1)^2 \De^2}.
\ee
Similarly, for the decomposition of $\mathcal{G}^{\cN=2 | JJ ; \phi \phi^{\dagger}}$ we expect
\be
\mathcal{G}^{\cN=2 | JJ ; \phi \phi^{\dagger}}_{\De,\ell,\text{even}} &=& \mathcal{G}^{\cN=1 | JJ ; \phi\phi^{\dagger}}_{\De,\ell,\text{even}} + \tl{N}(\De,\ell) \mathcal{G}^{\cN=1 | JJ ; \phi\phi^{\dagger}}_{\De+1,\ell - 1,\text{odd}} + \tl{J}(\De,\ell) \mathcal{G}^{\cN=1 | JJ ; \phi\phi^{\dagger}}_{\De+1,\ell + 1,\text{odd}} \nn \\
&& + \tl{D}(\De,\ell) \mathcal{G}^{\cN=1 | JJ ; \phi\phi^{\dagger}}_{\De+2,\ell,\text{even}},
\ee
with no contribution from (\ref{eq:newblock}) because these operators cannot appear in $\phi \times \phi^{\dagger}$. Matching to Eqs.~(\ref{eq:decomp}) and (\ref{eq:As}) gives
\be
\tl{N}(\De,\ell) &=& \frac{(\ell-1)(\De-\ell)(\De+\ell)}{8(\ell+1)(\De-\ell-1)},\\
\tl{J}(\De,\ell) &=& -\frac{(\De+\ell+2)^2}{8(\De+\ell+1)},\\
\tl{D}(\De,\ell) &=&-\frac{(\De+2)(\De-\ell)(\De+\ell+2)}{16\De(\De-\ell-1)(\De+\ell+1)}.
\ee
The existence of this decomposition provides a highly nontrivial check on the superconformal block results derived using our methods.

\section{Summary and Outlook}
\label{sec:summary}

We have computed the superconformal block $\mathcal{G}_{\Delta,\ell}^{\mathcal{N}=1 | A_1 A_2^{\dagger} ; B_1 B_2^{\dagger}}$, where $\cA_{1,2}$ and $\cB_{1,2}$ are scalar superconformal primaries with general dimensions and R-charge, and the exchanged operator is R-charge neutral. When $\cA_{1,2}$ and $\cB_{1,2}$ are chiral, we reproduce the known result for superconformal blocks in the chiral-antichiral channel. Similarly, when $\cA_{1,2}$ are global symmetry currents and $\cB_{1,2}$ are either global symmetry currents or chiral operators, we obtain expressions for $\mathcal{G}_{\Delta,\ell}^{\mathcal{N}=1 | JJ; JJ}$ and $\mathcal{G}_{\Delta,\ell}^{\mathcal{N}=1 | JJ; \phi\phi^\dagger}$.

There are many future directions to explore. Most immediately, the blocks we have computed provide new atomic ingredients to continue the study of $\cN=1$ SCFTs using the numerical bootstrap. We also hope to apply supershadow methods to SCFTs with $\cN>1$; for instance, the methods may be well-suited to study four-point functions of $\cN=2$ supercurrents. Regarding the supershadow formalism in general, it would be interesting to develop further machinery that allows us to: (i) compute shadow integrals in a manifestly superconformally covariant way, rather than just a conformally covariant way as we have done in this work; and (ii) consider superconformal blocks where the exchanged primaries have nonzero R-charge. We hope to explore these various directions in the near future.

\section*{Acknowledgements}

We are grateful to J. Kaplan and L. Fitzpatrick for collaboration and many discussions at the early stages of this work. We also thank C. Beem, M. Berkooz, J. Fortin, W. Goldberger, K. Intriligator, L. Rastelli, D. Skinner, A. Stergiou, R. Yacoby, and A. Zait for discussions and correspondence. We additionally thank the organizers of the ``Back to the Bootstrap 3" workshop at CERN for facilitating discussions related to this work. DP and DSD thank the KITP for its hospitality during the completion of this work. This research was supported in part by the National Science Foundation under Grant No.~NSF PHY11-25915. DSD was supported by DOE grant DE-SC0009988.  ZUK was supported by DOE grant 9500302471.

\newpage
\appendix

\section{Embedding-Space Derivatives and Integrals}
\label{app:Derivs}

In this appendix, we present some additional details of the calculation
peformed in Section~\ref{sec:PartialWave}. We describe several properties
of the quantities $\cN_{\ell}$, $D_{\ell}$, and $z$ that are useful
for computing the $\partial_{\bar{0}}^{2}$ derivative in Eq.~(\ref{W2}).
We present the results of this derivative acting on
various terms, and evaluate the relevant conformal
integrals. 
\subsection{$\cN_{\ell}$ as a Gegenbauer Polynomial}

One can show that the quantity $\cN_{\ell}$ in Eq.~(\ref{NL}) satisfies
a recursion relation that identifies it as a Gegenbauer polynomial,
\begin{equation}
\cN_{\ell}\equiv\left(\bar{\cS}1\bar{2}\cS\right)^{\ell}\overleftrightarrow{\mathcal{D}_{\ell}}\left(\bar{\cT}3\bar{4}\cT\right)^{\ell}=\left(-1\right)^{\ell}s^{\frac{\ell}{2}}C_{\ell}^{(1)}(t)\label{NL2},
\end{equation}
\noindent where $C_{\ell}^{(\lambda)}(x)$ are the Gegenbauer polynomials
and 
\begin{equation}
t\equiv\frac{\langle \bar{2}1\bar{0}3\bar{4}0\rangle }{2\sqrt{s}},\qquad\qquad s\equiv\frac{1}{2^{6}}\langle \bar{0}1\rangle \langle \bar{2}0\rangle \langle \bar{0}3\rangle \langle \bar{4}0\rangle \langle \bar{2}1\rangle \langle \bar{4}3\rangle .\label{s,t}
\end{equation}
\noindent Recall our notation,
\begin{equation}
\overleftrightarrow{\mathcal{D}_{\ell}}\equiv\frac{1}{\ell!^{4}}\left(\partial_{\cS}0\partial_{\cT}\right)^{\ell}\left(\partial_{\bar{\cS}}\bar{0}\partial_{\bar{\cT}}\right)^{\ell}.\label{Deriv2}
\end{equation}
One can also write $\cN_{\ell}$ as 
\begin{equation}
\cN_{\ell}=\frac{1}{\ell!^{2}}\left(\partial_{\cS}0\partial_{\cT}\right)^{\ell}\left(\cS\bar{2}1\bar{0}3\bar{4}\cT\right)^{\ell} .
\label{NLalt}
\end{equation}
This expression follows from Eq.~(\ref{NL2}) after acting all the $\partial_{\bar{\cS},\bar{\cT}}$ 
derivatives. 

When $\theta_{\mathrm{ext}}=0$, an $SU(2,2|1)$ trace reduces to an $SU(2,2)$ trace of the six-dimensional ``sigma" matrices $\Gamma^{m},\tl{\Gamma}^{m}$. 
In particular, $\cN_\ell(\cX_0,\bar{\cX}_0)$ reduces to a function of $(X_0,X_{\bar{0}})$. In this appendix, to be explicit, we define
\begin{equation}
N_\ell \left(X_0,X_{\bar{0}}\right) \equiv \left.\cN_{\ell}\right|_{\theta_{\mathrm{ext}}=0}.
\end{equation} 
It is given by the same expressions as Eqs.~(\ref{NL2}) and (\ref{NLalt}), except that all auxiliary twistors and coordinates are reduced to their bosonic twistor parts. We write this simply as, for instance, $\left(\cS\bar{2}1\bar{0}3\bar{4}\cT\right)\rightarrow\left(S\bar{2}1\bar{0}3\bar{4}T\right)$, etc. 

With $\theta_{\mathrm{ext}}=0$, but prior to the $\partial_{\bar{0}}^{2}$ differentiation, $\bar{0}$ and $0$ are considered to be independent. Afterwards, though, we identify $\bar{0}=0$, in which case
\begin{eqnarray}
t&\longrightarrow& t_{0}\equiv-\frac{X_{13}X_{20}X_{40}}{2\sqrt{X_{10}X_{20}X_{30}X_{40}X_{12}X_{34}}}-\left(1\leftrightarrow2\right)-\left(3\leftrightarrow4\right), \label{t0} \\
s&\longrightarrow& s_{0}\equiv\frac{1}{2^{12}}X_{10}X_{20}X_{30}X_{40}X_{12}X_{34}.\label{s0}
\end{eqnarray}
\noindent Recalling that
\begin{equation}
\frac{1}{D_{\ell}}\equiv\frac{1}{\left(X_{10}X_{1\bar{0}}X_{20}X_{2\bar{0}}\right)^{\frac14(\De+\ell)}\left(X_{30}X_{3\bar{0}}X_{40}X_{4\bar{0}}\right)^{-\frac14(\De-\ell-2)}} ,
\end{equation}
\noindent we therefore have that
\begin{equation}
\left.\frac{N_{\ell}}{D_{\ell}}\right|_{\bar{0}=0}=a_{\ell}\left(X_{12}X_{34}\right)^{\frac{\ell}{2}}\frac{(-1)^\ell C_{\ell}^{(1)}(t_{0})}{\left(X_{10}X_{20}\right)^{\frac{\De}{2}}\left(X_{30}X_{40}\right)^{1-\frac{\De}{2}}}\label{NL/DL} ,
\end{equation}
\noindent where
\begin{equation}
a_{\ell}\equiv 2^{-6\ell}.
\label{aL}
\end{equation}
\subsection{Symmetries of $N_{\ell}$, $D_{\ell}$, and $z$}
First, consider 
\begin{equation}
N_{\ell}=(\bar{S}1\bar{2}S)^{\ell}\overleftrightarrow{\mathcal{D}_{\ell}}(\bar{T}3\bar{4}T)^{\ell}=\frac{1}{\ell!^{2}}(\partial_{S}0\partial_{T})^{\ell}(S\bar{2}1\bar{0}3\bar{4}T)^{\ell} .
\label{nL}
\end{equation}
If we identify $\bar{0}$ with $0$, then 
\begin{equation}
S\bar{2}1\bar{0}3\bar{4}T=\frac{1}{4}X_{10}S\bar{2}3\bar{4}T-(1\leftrightarrow2)=\frac{1}{4}X_{30}S\bar{2}1\bar{4}T-(3\leftrightarrow4)\label{Symm2} .
\end{equation}
\noindent These expressions can be derived by using the Clifford algebra
of the sigma matrices to commute $\bar{0}$ to the left or right and
then using $S\bar{0}=\bar{0}T=0$ (see Eq.~(\ref{transverse})). It follows that $S\bar{2}1\bar{0}3\bar{4}T$
is antisymmetric in $1\leftrightarrow2$ and $3\leftrightarrow4$,
so $N_{\ell}$ is either symmetric or antisymmetric in these exchanges
depending on the parity of $\ell$. This also follows from our expression
for $N_{\ell}$ as a Gegenbauer polynomial. 

Since $\partial_{\bar{0}}^{2}N_{\ell}\propto(\partial_{\bar{S}}\bar{\Gamma}^{m}\partial_{\bar{T}})(\partial_{\bar{S}}\bar{\Gamma}^{m}\partial_{\bar{T}})\propto\epsilon^{\alpha\beta\gamma\delta}\partial_{\bar{S}\alpha}\partial_{\bar{T}\beta}\partial_{\bar{S}\gamma}\partial_{\bar{T}\delta}=0$,
we have the important result that 
\begin{equation}
\partial_{\bar{0}}^{2}N_{\ell}=0
\end{equation}

Meanwhile, $D_{\ell}$ is always symmetric under the exchanges $1\leftrightarrow2$
and/or $3\leftrightarrow4$. Its derivatives are
\begin{eqnarray}
\partial_{\bar{0}}^{m}\frac{1}{D_{\ell}}&=&\frac{1}{D_{\ell}}\left[\left(\frac{\De+\ell}{2}\right)\left(\frac{X_{1}^{m}}{X_{1\bar{0}}}+\frac{X_{2}^{m}}{X_{2\bar{0}}}\right)-\left(\frac{\De-\ell-2}{2}\right)\left(\frac{X_{3}^{m}}{X_{3\bar{0}}}+\frac{X_{4}^{m}}{X_{4\bar{0}}}\right)\right], \\
\partial_{\bar{0}}^{2}\frac{1}{D_{\ell}}&=&-\frac{1}{D_{\ell}}\left[\left(\frac{\De+\ell}{2}\right)^{2}\frac{X_{12}}{X_{1\bar{0}}X_{2\bar{0}}}+\left(\frac{\De-\ell-2}{2}\right)^{2}\frac{X_{34}}{X_{3\bar{0}}X_{4\bar{0}}}\right. \nonumber \\
&&\hspace{20mm} \left. -\left(\frac{\De+\ell}{2}\right)\left(\frac{\De-\ell-2}{2}\right)\left(\frac{X_{13}}{X_{1\bar{0}}X_{3\bar{0}}}+\mathrm{evenperms}\right)\right],
\end{eqnarray}
\noindent where\footnote{Recall our notation: $\left(3\leftrightarrow4\right)$ acts on both the first term and the $\left(1\leftrightarrow2\right)$ term.}
\begin{equation}
\mathrm{evenperms}=+\left(1\leftrightarrow2\right)+\left(3\leftrightarrow4\right).
\end{equation}

Finally, $z$ is antisymmetric under $1\leftrightarrow2$, while $\tl{z}$
is antisymmetric under $3\leftrightarrow4$. Its derivatives are
\begin{equation}
\partial_{\bar{0}}^{m}z=(z^{2}-1)\left[\frac{X_{1}^{m}}{X_{1\bar{0}}}-\frac{X_{2}^{m}}{X_{2\bar{0}}}\right],\qquad\partial_{\bar{0}}^{2}z=2z(z^{2}-1)\frac{X_{12}}{X_{1\bar{0}}X_{2\bar{0}}},
\end{equation}
\begin{equation}
\partial_{\bar{0}}^{m}z^{2}=2z(z^{2}-1)\left[\frac{X_{1}^{m}}{X_{1\bar{0}}}-\frac{X_{2}^{m}}{X_{2\bar{0}}}\right],\qquad\partial_{\bar{0}}^{2}z^{2}=(6z^{2}-2)(z^{2}-1)\frac{X_{12}}{X_{1\bar{0}}X_{2\bar{0}}}.
\end{equation}

\noindent Similar formulas hold for $\tl{z}$ with $1,2$ replaced
by $3,4$ respectively. In particular
\begin{equation}
\left.(\partial_{\bar{0}}z)\cdot(\partial_{\bar{0}}\tl{z})\right|_{\bar{0}=0}=-\frac{1}{2}\left[\frac{X_{13}}{X_{10}X_{30}}+\mathrm{oddperms}\right],
\end{equation}
\noindent where
\begin{equation}
\mathrm{oddperms}=-\left(1\leftrightarrow2\right)-\left(3\leftrightarrow4\right).
\end{equation}
\noindent An important point is that $z$ and $\tl{z}$ vanish
when $\theta_{\mathrm{ext}}=0$ and $\bar{0}$ is identified with $0$ (i.e.
without supersymmetry, there is no conformally-invariant cross-ratio
given three points). This simplifies derivatives involving these quantities,
because the derivatives must act to eliminate all $z$'s and $\tl{z}$'s.

\subsection{Results for $\partial_{\bar{0}}^{2}$ Derivatives}
Using the equations above, it follows that
\begin{eqnarray}
\left.\partial_{\bar{0}}^{2}\frac{z^{2}N_{\ell}}{D_{\ell}}\right|_{\bar{0}=0}&=&2\frac{X_{12}}{X_{10}X_{20}}\frac{N_{\ell}}{D_{\ell}}, \\
\left.\partial_{\bar{0}}^{2}\frac{z\tl{z}N_{\ell}}{D_{\ell}}\right|_{\bar{0}=0}&=&-\left[\frac{X_{13}}{X_{10}X_{30}}+\mathrm{oddperms}\right]\frac{N_{\ell}}{D_{\ell}}, \\
\left.\partial_{\bar{0}}^{2}\frac{zN_{\ell}}{D_{\ell}}\right|_{\bar{0}=0}&=&-\left(\frac{\De-\ell-2}{2}\right)\frac{N_{\ell}}{D_{\ell}}\left[\frac{X_{13}}{X_{10}X_{30}}-\frac{X_{23}}{X_{20}X_{30}}+\frac{X_{14}}{X_{10}X_{40}}-\frac{X_{24}}{X_{20}X_{40}}\right] \nonumber \\
&&  +\frac{1}{2}\frac{1}{D_{\ell}}\frac{\ell}{\ell!^{2}}\left(\partial_{S}0\partial_{T}\right)^{\ell}\left(S\bar{2}1\bar{0}3\bar{4}T\right)^{\ell-1}\left[\frac{X_{12}}{X_{10}X_{20}}X_{10}\left(S\bar{2}3\bar{4}T\right)\right], \\
\left.\partial_{\bar{0}}^{2}\frac{N_{\ell}}{D_{\ell}}\right|_{\bar{0}=0}&=&-\frac{N_{\ell}}{D_{\ell}}\left[\left(\frac{\De+\ell}{2}\right)^{2}\frac{X_{12}}{X_{10}X_{20}}+\left(\frac{\De-\ell-2}{2}\right)^{2}\frac{X_{34}}{X_{30}X_{40}} \right. \nonumber \\
&& \hspace{20mm}
\left.-\left(\frac{\De+\ell}{2}\right)\left(\frac{\De-\ell-2}{2}\right)\left(\frac{X_{13}}{X_{10}X_{30}}+\mathrm{evenperms}\right)\right] \nonumber \\
&& +\frac{1}{2}\frac{1}{D_{\ell}}\frac{\ell}{\ell!^{2}}\left(\partial_{S}0\partial_{T}\right)^{\ell}\left(S\bar{2}1\bar{0}3\bar{4}T\right)^{\ell-1}\left[\left(\frac{\De+\ell}{2}\right)\frac{X_{12}}{X_{10}X_{20}}\left(X_{10}S\bar{2}3\bar{4}T\right) \right. \nonumber \\
&& \hspace{20mm}
\left. -\left(\frac{\De-\ell-2}{2}\right)\frac{X_{34}}{X_{30}X_{40}}\left(X_{30}S\bar{2}1\bar{4}T\right)\right] .
\end{eqnarray}
\noindent Using these, one can derive the following derivative formulas
needed for Eq.~(\ref{W2})
\begin{eqnarray}
&& \hspace{-7mm}
4^{-\ell}X_{12}^{\ell}X_{34}^{\ell}\left.\partial_{\bar{0}}^{2}\left[z^{2}\left(S_{-}\right)^{\ell}\overleftrightarrow{\mathcal{D}_{\ell}}\left(T_{-}\right)^{\ell}\right]\right|_{\bar{0}=0}=2\frac{X_{12}}{X_{10}X_{20}}\frac{N_{\ell}}{D_{\ell}}, \\
&& \hspace{-7mm}
4^{-\ell}X_{12}^{\ell}X_{34}^{\ell}\left.\partial_{\bar{0}}^{2}\left[z\tl{z}\left(S_{-}\right)^{\ell}\overleftrightarrow{\mathcal{D}_{\ell}}\left(T_{-}\right)^{\ell}\right]\right|_{\bar{0}=0}=-\left[\frac{X_{13}}{X_{10}X_{30}}+\mathrm{oddperms}\right]\frac{N_{\ell}}{D_{\ell}}, \\
&& \hspace{-7mm}
4^{-\ell}X_{12}^{\ell}X_{34}^{\ell}\left.\partial_{\bar{0}}^{2}\left[z\left(S_{-}\right)^{\ell}\overleftrightarrow{\mathcal{D}_{\ell}}T_{+}\left(T_{-}\right)^{\ell-1}\right]\right|_{\bar{0}=0}=-\frac{1}{2^{6}}\frac{\left(\ell+1\right)}{\ell}\frac{N_{\ell-1}}{D_{\ell}}X_{12}X_{34}, \\
&& \hspace{-7mm}
4^{-\ell}X_{12}^{\ell}X_{34}^{\ell}\left.\partial_{\bar{0}}^{2}\left[S_{+}\left(S_{-}\right)^{\ell-1}\overleftrightarrow{\mathcal{D}_{\ell}}T_{+}\left(T_{-}\right)^{\ell-1}\right]\right|_{\bar{0}=0}=-\frac{1}{2^{6}}\frac{\left(\ell+1\right)^{2}}{\ell^{2}}X_{12}X_{34}\frac{N_{\ell-1}}{D_{\ell}}, \\
&& \hspace{-7mm}
4^{-\ell}X_{12}^{\ell}X_{34}^{\ell}\left.\partial_{\bar{0}}^{2}\left[\left(S_{-}\right)^{\ell}\overleftrightarrow{\mathcal{D}_{\ell}}\left(T_{-}\right)^{\ell}\right]\right|_{\bar{0}=0}= 
-\frac{N_{\ell}}{D_{\ell}}\left[\left(\frac{\De+\ell}{2}\right)\left(\frac{\De-\ell}{2}\right)\frac{X_{12}}{X_{10}X_{20}} \right. \nonumber\\
&&\hspace{59mm} \left.+\left(\frac{\De-\ell-2}{2}\right)\left(\frac{\De+\ell-2}{2}\right)\frac{X_{34}}{X_{30}X_{40}} \right. \nonumber\\
&&\hspace{59mm} \left.-\left(\frac{\De+\ell}{2}\right)\left(\frac{\De-\ell-2}{2}\right)\left(\frac{X_{13}}{X_{10}X_{30}}+\mathrm{evenperms}\right)\right] . \nonumber \\
\end{eqnarray}

\subsection{Conformal Integrals}
Once the differentiation is done in Eq.~(\ref{W2}), the final step
is to evaluate the resulting conformal integrals. The relevant formulas, with $a_\ell$ given by Eq.~(\ref{aL}), are 
\begin{eqnarray}
\int D^{4}X_{0}\left.\frac{X_{12}}{X_{10}X_{20}}\frac{N_{\ell}}{D_{\ell}}\right|_{\bar{0}=0}&=&a_{\ell}\,\frac{\xi_{\De+2,2-\De,0,\ell}}{X_{12}^{\frac12(\De-\ell)}X_{34}^{-\frac12(\De+\ell-2)}}\,g_{\De+2,\ell}^{0,0}(u,v) , \\
\int D^{4}X_{0}\left.\frac{X_{34}}{X_{30}X_{40}}\frac{N_{\ell}}{D_{\ell}}\right|_{\bar{0}=0}&=&a_{\ell}\,\frac{\xi_{\De,4-\De,0,\ell}}{X_{12}^{\frac12(\De-\ell)}X_{34}^{-\frac12(\De+\ell-2)}}\,g_{\De,\ell}^{0,0}(u,v) , \\
\int D^{4}X_{0}\left.\frac{N_{\ell-1}}{D_{\ell}}\right|_{\bar{0}=0}&=&a_{\ell-1}\,\frac{\xi_{\De+1,3-\De,0,\ell-1}}{X_{12}^{\frac12(\De-\ell)+1}X_{34}^{-\frac12(\De+\ell-2)+1}}\,g_{\De+1,\ell-1}^{0,0}(u,v) ,
\end{eqnarray}
\begin{eqnarray}
\int D^{4}X_{0}\left.\left[\frac{X_{13}}{X_{10}X_{30}}+\mathrm{evenperms}\right]\frac{N_{\ell}}{D_{\ell}}\right|_{\bar{0}=0}&=& \nonumber\\
&&\hspace{-60mm}
a_{\ell}\,\frac{\xi_{\De+1,3-\De,1,\ell}}{X_{12}^{\frac12(\De-\ell)}X_{34}^{-\frac12(\De+\ell-2)}}\left[4g_{\De,\ell}^{0,0}+\frac{\left(\De+\ell\right)\left(\De-\ell-2\right)}{4\left(\De+\ell+1\right)\left(\De-\ell-1\right)}g_{\De+2,\ell}^{0,0}\right] , \\
\int D^{4}X_{0}\left.\left[\frac{X_{13}}{X_{10}X_{30}}+\mathrm{oddperms}\right]\frac{N_{\ell}}{D_{\ell}}\right|_{\bar{0}=0} &=& \nonumber \\
&&\hspace{-60mm}
a_{\ell}\,\frac{\xi_{\De+1,3-\De,1,\ell}}{X_{12}^{\frac12(\De-\ell)}X_{34}^{-\frac12(\De+\ell-2)}}\left[\frac{\left(\De+\ell\right)}{\left(\De+\ell+1\right)}g_{\De+1,\ell+1}^{0,0}+\frac{\left(\De-\ell-2\right)}{\left(\De-\ell-1\right)}g_{\De+1,\ell-1}^{0,0}\right] .\nn\\
\end{eqnarray}

\noindent The first three equations follow from Eq.~(\ref{NL/DL})
and a direct application of Eq.~(\ref{ConfInt}). For the latter two
integrals, one additionally needs the following relations between
the conformal blocks,
\begin{eqnarray}
g_{\Delta,\ell}^{-\Delta_{12},-\Delta_{34}}&=&v^{\frac{1}{2}\left(\Delta_{34}-\Delta_{12}\right)}g_{\Delta,\ell}^{\Delta_{12},\Delta_{34}} , \\
u^{-\frac{1}{2}}g_{\De+1,\ell}^{1,1}+u^{-\frac{1}{2}}g_{\De+1,\ell}^{1,-1}&=&2g_{\De,\ell}^{0,0}+\frac{\left(\De+\ell\right)\left(\De-\ell-2\right)}{8\left(\De+\ell+1\right)\left(\De-\ell-1\right)}g_{\De+2,\ell}^{0,0} , \\
u^{-\frac{1}{2}}g_{\De+1,\ell}^{1,1}-u^{-\frac{1}{2}}g_{\De+1,\ell}^{1,-1}&=&\frac{\left(\De+\ell\right)}{2\left(\De+\ell+1\right)}g_{\De+1,\ell+1}^{0,0}+\frac{\left(\De-\ell-2\right)}{2\left(\De-\ell-1\right)}g_{\De+1,\ell-1}^{0,0} .
\end{eqnarray}

\section{Shadow 3-Point Function Coefficients}
\label{app:shadowtransformation}
In this appendix we will show how one can derive the transformation matrices in Eq.~(\ref{eq:ShadowTransformation}). Because 3-point function structures can only mix with structures with the same symmetry properties, the linear transformation $\l^{(i)}_{\cB_1\cB_2^\dag\tl{\cO}} = \cM(\De,\ell)^i_j \l^{(j)}_{\cB_1\cB_2^\dag\cO}$ must be block diagonal
\be
\left( \begin{array}{c}
\lambda_{\cB_1\cB_2^\dag\tl{\cO}}^{(0)} \\
\lambda_{\cB_1\cB_2^\dag\tl{\cO}}^{(2)} \\
\lambda_{\cB_1\cB_2^\dag\tl{\cO}}^{(1)} \\
\lambda_{\cB_1\cB_2^\dag\tl{\cO}}^{(3)} 
\end{array} \right) 
& = & 
\left( \begin{array}{cccc}
A & B & 0 & 0 \\
C & D & 0 & 0 \\
0 & 0 & E & F \\
0 & 0 & G & H 
\end{array} \right)
\left( \begin{array}{c}
\lambda_{\cB_1\cB_2^\dag\cO}^{(0)} \\
\lambda_{\cB_1\cB_2^\dag\cO}^{(2)} \\
\lambda_{\cB_1\cB_2^\dag\cO}^{(1)} \\
\lambda_{\cB_1\cB_2^\dag\cO}^{(3)} 
\end{array} \right) .
\ee
Taking $\cB_1 = \cB_2 = \Phi$ to be chiral and using Eq.~(\ref{eq:Chiral3PF2}), we have the constraint
\be
\left( \begin{array}{c}
 1 \\
\frac{1}{8}(2-\De+\ell)(2-\De-\ell) \\
-\frac{1}{2}(2-\De+\ell) \\
\ell
\end{array} \right)
& \propto & 
\left( \begin{array}{cccc}
A & B & 0 & 0 \\
C & D & 0 & 0 \\
0 & 0 & E & F \\
0 & 0 & G & H 
\end{array} \right)
\left( \begin{array}{c}
 1 \\
\frac{1}{8}(\De+\ell)(\De-\ell) \\
-\frac{1}{2}(\De+\ell) \\
\ell
\end{array} \right) .\nn\\
\ee
On the other hand, taking $\cB_i = \cJ_i$ to be a conserved current and using Eqs.~(\ref{eq:Conservation}) gives the constraints
\be
\left( \begin{array}{c}
1 \\
\frac{1}{8}(2+\ell+\De)(2-\De+\ell)\\
\end{array} \right)
& \propto & 
\left( \begin{array}{cc}
A & B \\
C & D \\
\end{array} \right)
\left( \begin{array}{c}
1 \\
\frac{1}{8}(4+\ell-\De)(\De+\ell) \\
\end{array} \right) ,\nn \\
\left( \begin{array}{c}
\frac{(2-\De+\ell)}{2\De}  \\
1
\end{array} \right)
& \propto & 
\left( \begin{array}{cc}
E & F \\
G & H 
\end{array} \right)
\left( \begin{array}{c}
-\frac{(\De+\ell)}{2(\De-2)}  \\
 1
\end{array} \right) .
\ee
The above equations so far give 5 constraints on 8 unknowns. 

The remaining freedom can be fixed by requiring that the superconformal partial waves are consistent with unitarity, i.e. that in reflection positive configurations the coefficients of individual conformal blocks have positive coefficients. Concretely, the computations described in Section~\ref{sec:PartialWave} and Appendix~\ref{app:Derivs} give the result
\be
\mathcal{G}_{\De,\ell}^{\mathcal{N}=1 | A_1 A_2^\dag ; B_1 B_2^\dag}
&\propto& \nn \\
&& \hspace{-20mm}
\lambda_{\cA_1\cA_2^\dag\cO}^{(0)}\left[\lambda_{\cB_1\cB_2^\dag\tl{\cO}}^{(2)}+\lambda_{\cB_1\cB_2^\dag\tl{\cO}}^{(0)}\frac{(\Delta-\ell-2)^{2}}{8(\Delta-1)}\right]g_{\Delta,\ell} \nn \\
&& \hspace{-20mm}
+\lambda_{\cA_1\cA_2^\dag\cO}^{(1)}\lambda_{\cB_1\cB_2^\dag\tl{\cO}}^{(1)}\frac{(\Delta-2)(\Delta+\ell)}{8(\Delta-1)(\Delta+\ell+1)}g_{\Delta+1,\ell+1}\nn\\
&& \hspace{-20mm}
+\left[\lambda_{\cA_1\cA_2^\dag\cO}^{(1)}+\frac{\ell+1}{\ell}\lambda_{\cA_1\cA_2^\dag\cO}^{(3)}\right]\left[\lambda_{\cB_1\cB_2^\dag\tl{\cO}}^{(1)}+\frac{\ell+1}{\ell}\lambda_{\cB_1\cB_2^\dag\tl{\cO}}^{(3)}\right]\frac{(\Delta-2)(\Delta-\ell-2)}{8\left(\Delta-1\right)\left(\Delta-\ell-1\right)}g_{\Delta+1,\ell-1}\nn\\
&& \hspace{-20mm}
+\left[\lambda_{\cA_1\cA_2^\dag\cO}^{(2)}-\lambda_{\cA_1\cA_2^\dag\cO}^{(0)}\frac{(\Delta+\ell)^{2}}{8(\Delta-1)}\right]\lambda_{\cB_1\cB_2^\dag\tl{\cO}}^{(0)}\frac{(\Delta-2)(\Delta+\ell)(\Delta-\ell-2)}{16\Delta(\Delta+\ell+1)(\Delta-\ell-1)}g_{\Delta+2,\ell} .
\ee
In the reflection positive configuration $\cB_{1,2} = \cA_{2,1}$, each of the conformal block coefficients must be positive. This implies that
\be
\left[\lambda_{\cB_1\cB_2^\dag\tl{\cO}}^{(2)}+\lambda_{\cB_1\cB_2^\dag\tl{\cO}}^{(0)}\frac{(\Delta-\ell-2)^{2}}{8(\Delta-1)}\right] &\propto& \l^{(0)}_{\cB_1\cB_2^\dag\cO}, \\
\lambda_{\cB_1\cB_2^\dag\tl{\cO}}^{(1)} &\propto& \lambda_{\cB_1\cB_2^\dag\cO}^{(1)}, \\
\left[\lambda_{\cB_1\cB_2^\dag\tl{\cO}}^{(1)}+\frac{\ell+1}{\ell}\lambda_{\cB_1\cB_2^\dag\tl{\cO}}^{(3)}\right] &\propto& \left[\lambda_{\cB_1\cB_2^\dag\cO}^{(1)}+\frac{\ell+1}{\ell}\lambda_{\cB_1\cB_2^\dag\cO}^{(3)}\right], \\
\lambda_{\cB_1\cB_2^\dag\tl{\cO}}^{(0)} &\propto& \left[\lambda_{\cB_1\cB_2^\dag\cO}^{(2)}-\lambda_{\cB_1\cB_2^\dag\cO}^{(0)}\frac{(\Delta+\ell)^{2}}{8(\Delta-1)}\right],
\ee
which imposes the additional constraints
\be
A + \frac{(\De+\ell)^2}{8(\De-1)} B = D+\frac{(\De-\ell-2)^2}{8(\De-1)} B = H - E - \frac{\ell+1}{\ell} G = F = 0 .
\ee
This fixes the transformation matrix up to an overall constant
\be
\left( \begin{array}{cccc}
A & B & 0 & 0 \\
C & D & 0 & 0 \\
0 & 0 & E & F \\
0 & 0 & G & H 
\end{array} \right) 
&=&
H \left( \begin{array}{cccc}
\frac{1}{\Delta} & -\frac{8\left(\Delta-1\right)}{\Delta\left(\Delta+\ell\right)^{2}} & 0 & 0 \\
- \frac{\left(\Delta-1\right)\left(\Delta-\ell-2\right)^{2}}{8\Delta} & \frac{\left(\Delta-\ell-2\right)^{2}}{\Delta\left(\Delta+\ell\right)^{2}} & 0 & 0 \\
0 & 0 & \frac{\left(\Delta-\ell-2\right)^{2}}{\left(\Delta+\ell\right)^{2}} & 0 \\
0 & 0 & \frac{4\ell\left(\Delta-1\right)}{\left(\Delta+\ell\right)^{2}} & 1
\end{array} \right),
\ee
where $H$ depends on the overall normalization of the shadow transformation. 
A convenient choice is $H=1$, which corresponds to defining the shadow transformation $\cM(\De,\ell)$ so that when applied twice it gives the identity $\cM(\De,\ell) \cdot \cM(2-\De,\ell) = \mathbf{1}_{4\x4}$.

\newpage
\bibliography{Biblio}{}

\providecommand{\href}[2]{#2}\begingroup\raggedright\begin{thebibliography}{10}

\bibitem{Polyakov:1974gs}
A.~M. Polyakov, ``{Nonhamiltonian approach to conformal quantum field
  theory},''
{\em Zh. Eksp. Teor. Fiz.} {\bfseries 66} (1974) 23--42.

\bibitem{Rattazzi:2008pe}
R.~Rattazzi, V.~S. Rychkov, E.~Tonni, and A.~Vichi, ``{Bounding scalar operator
  dimensions in 4D CFT},''
  \href{http://dx.doi.org/10.1088/1126-6708/2008/12/031}{{\em JHEP} {\bfseries
  12} (2008) 031},
\href{http://arxiv.org/abs/0807.0004}{{\ttfamily arXiv:0807.0004 [hep-th]}}.

\bibitem{Rychkov:2009ij}
V.~S. Rychkov and A.~Vichi, ``{Universal Constraints on Conformal Operator
  Dimensions},'' \href{http://dx.doi.org/10.1103/PhysRevD.80.045006}{{\em Phys.
  Rev.} {\bfseries D80} (2009) 045006},
\href{http://arxiv.org/abs/0905.2211}{{\ttfamily arXiv:0905.2211 [hep-th]}}.

\bibitem{Caracciolo:2009bx}
F.~Caracciolo and V.~S. Rychkov, ``{Rigorous Limits on the Interaction Strength
  in Quantum Field Theory},''
  \href{http://dx.doi.org/10.1103/PhysRevD.81.085037}{{\em Phys. Rev.}
  {\bfseries D81} (2010) 085037},
\href{http://arxiv.org/abs/0912.2726}{{\ttfamily arXiv:0912.2726 [hep-th]}}.

\bibitem{Poland:2010wg}
D.~Poland and D.~Simmons-Duffin, ``{Bounds on 4D Conformal and Superconformal
  Field Theories},'' \href{http://dx.doi.org/10.1007/JHEP05(2011)017}{{\em
  JHEP} {\bfseries 1105} (2011) 017},
\href{http://arxiv.org/abs/1009.2087}{{\ttfamily arXiv:1009.2087 [hep-th]}}.

\bibitem{Rattazzi:2010gj}
R.~Rattazzi, S.~Rychkov, and A.~Vichi, ``{Central Charge Bounds in 4D Conformal
  Field Theory},'' \href{http://dx.doi.org/10.1103/PhysRevD.83.046011}{{\em
  Phys. Rev.} {\bfseries D83} (2011) 046011},
\href{http://arxiv.org/abs/1009.2725}{{\ttfamily arXiv:1009.2725 [hep-th]}}.

\bibitem{Rattazzi:2010yc}
R.~Rattazzi, S.~Rychkov, and A.~Vichi, ``{Bounds in 4D Conformal Field Theories
  with Global Symmetry},''
  \href{http://dx.doi.org/10.1088/1751-8113/44/3/035402}{{\em J. Phys.}
  {\bfseries A44} (2011) 035402},
\href{http://arxiv.org/abs/1009.5985}{{\ttfamily arXiv:1009.5985 [hep-th]}}.

\bibitem{Vichi:2011ux}
A.~Vichi, ``{Improved bounds for CFT's with global symmetries},''
  \href{http://dx.doi.org/10.1007/JHEP01(2012)162}{{\em JHEP} {\bfseries 1201}
  (2012) 162}, \href{http://arxiv.org/abs/1106.4037}{{\ttfamily arXiv:1106.4037
  [hep-th]}}.
24 pages, 6 figures.

\bibitem{Poland:2011ey}
D.~Poland, D.~Simmons-Duffin, and A.~Vichi, ``{Carving Out the Space of 4D
  CFTs},'' \href{http://dx.doi.org/10.1007/JHEP05(2012)110}{{\em JHEP}
  {\bfseries 1205} (2012) 110},
\href{http://arxiv.org/abs/1109.5176}{{\ttfamily arXiv:1109.5176 [hep-th]}}.

\bibitem{Rychkov:2011et}
S.~Rychkov, ``{Conformal Bootstrap in Three Dimensions?},''
\href{http://arxiv.org/abs/1111.2115}{{\ttfamily arXiv:1111.2115 [hep-th]}}.

\bibitem{ElShowk:2012ht}
S.~El-Showk, M.~F. Paulos, D.~Poland, S.~Rychkov, D.~Simmons-Duffin, and
  A.~Vichi, ``{Solving the 3D Ising Model with the Conformal Bootstrap},''
  \href{http://dx.doi.org/10.1103/PhysRevD.86.025022}{{\em Phys.Rev.}
  {\bfseries D86} (2012) 025022},
\href{http://arxiv.org/abs/1203.6064}{{\ttfamily arXiv:1203.6064 [hep-th]}}.

\bibitem{Liendo:2012hy}
P.~Liendo, L.~Rastelli, and B.~C. van Rees, ``{The Bootstrap Program for
  Boundary $CFT_d$},'' \href{http://dx.doi.org/10.1007/JHEP07(2013)113}{{\em
  JHEP} {\bfseries 1307} (2013) 113},
\href{http://arxiv.org/abs/1210.4258}{{\ttfamily arXiv:1210.4258 [hep-th]}}.

\bibitem{ElShowk:2012hu}
S.~El-Showk and M.~F. Paulos, ``{Bootstrapping Conformal Field Theories with
  the Extremal Functional Method},''
\href{http://arxiv.org/abs/1211.2810}{{\ttfamily arXiv:1211.2810 [hep-th]}}.

\bibitem{Fitzpatrick:2012yx}
A.~L. Fitzpatrick, J.~Kaplan, D.~Poland, and D.~Simmons-Duffin, ``{The Analytic
  Bootstrap and AdS Superhorizon Locality},''
  \href{http://dx.doi.org/10.1007/JHEP12(2013)004}{{\em JHEP} {\bfseries 1312}
  (2013) 004},
\href{http://arxiv.org/abs/1212.3616}{{\ttfamily arXiv:1212.3616 [hep-th]}}.

\bibitem{Komargodski:2012ek}
Z.~Komargodski and A.~Zhiboedov, ``{Convexity and Liberation at Large Spin},''
  \href{http://dx.doi.org/10.1007/JHEP11(2013)140}{{\em JHEP} {\bfseries 1311}
  (2013) 140},
\href{http://arxiv.org/abs/1212.4103}{{\ttfamily arXiv:1212.4103 [hep-th]}}.

\bibitem{Beem:2013qxa}
C.~Beem, L.~Rastelli, and B.~C. van Rees, ``{The N=4 Superconformal
  Bootstrap},'' \href{http://dx.doi.org/10.1103/PhysRevLett.111.071601}{{\em
  Phys.Rev.Lett.} {\bfseries 111} (2013) 071601},
\href{http://arxiv.org/abs/1304.1803}{{\ttfamily arXiv:1304.1803 [hep-th]}}.

\bibitem{Kos:2013tga}
F.~Kos, D.~Poland, and D.~Simmons-Duffin, ``{Bootstrapping the O(N) Vector
  Models},''
\href{http://arxiv.org/abs/1307.6856}{{\ttfamily arXiv:1307.6856}}.

\bibitem{Gliozzi:2013ysa}
F.~Gliozzi, ``{More constraining conformal bootstrap},''
  \href{http://dx.doi.org/10.1103/PhysRevLett.111.161602}{{\em Phys.Rev.Lett.}
  {\bfseries 111} (2013) 161602},
\href{http://arxiv.org/abs/1307.3111}{{\ttfamily arXiv:1307.3111}}.

\bibitem{El-Showk:2013nia}
S.~El-Showk, M.~F. Paulos, D.~Poland, S.~Rychkov, D.~Simmons-Duffin, and
  A.~Vichi, ``{Conformal Field Theories in Fractional Dimensions},'' {\em Phys.
  Rev. Lett., to appear} (2013) ,
\href{http://arxiv.org/abs/1309.5089}{{\ttfamily arXiv:1309.5089 [hep-th]}}.

\bibitem{Alday:2013opa}
L.~F. Alday and A.~Bissi, ``{The superconformal bootstrap for structure
  constants},''
\href{http://arxiv.org/abs/1310.3757}{{\ttfamily arXiv:1310.3757 [hep-th]}}.

\bibitem{Gaiotto:2013nva}
D.~Gaiotto, D.~Mazac, and M.~F. Paulos, ``{Bootstrapping the 3d Ising twist
  defect},''
\href{http://arxiv.org/abs/1310.5078}{{\ttfamily arXiv:1310.5078 [hep-th]}}.

\bibitem{Bashkirov:2013vya}
D.~Bashkirov, ``{Bootstrapping the $\cN=1$ SCFT in three dimensions},''
\href{http://arxiv.org/abs/1310.8255}{{\ttfamily arXiv:1310.8255 [hep-th]}}.

\bibitem{Beem:2013sza}
C.~Beem, M.~Lemos, P.~Liendo, W.~Peelaers, L.~Rastelli, and B.~van Rees,
  ``{Infinite Chiral Symmetry in Four Dimensions},''
\href{http://arxiv.org/abs/1312.5344}{{\ttfamily arXiv:1312.5344 [hep-th]}}.

\bibitem{Berkooz:2014yda}
M.~Berkooz, R.~Yacoby, and A.~Zait, ``{Bounds on $\mathcal{N}=1$ Superconformal
  Theories with Global Symmetries},''
\href{http://arxiv.org/abs/1402.6068}{{\ttfamily arXiv:1402.6068 [hep-th]}}.

\bibitem{El-Showk:2014dwa}
S.~El-Showk, M.~F. Paulos, D.~Poland, S.~Rychkov, D.~Simmons-Duffin, {\em
  et~al.}, ``{Solving the 3d Ising Model with the Conformal Bootstrap II.
  c-Minimization and Precise Critical Exponents},''
\href{http://arxiv.org/abs/1403.4545}{{\ttfamily arXiv:1403.4545 [hep-th]}}.

\bibitem{Gliozzi:2014jsa}
F.~Gliozzi and A.~Rago, ``{Critical exponents of the 3d Ising and related
  models from Conformal Bootstrap},''
\href{http://arxiv.org/abs/1403.6003}{{\ttfamily arXiv:1403.6003 [hep-th]}}.

\bibitem{Fitzpatrick:2014vua}
A.~L. Fitzpatrick, J.~Kaplan, and M.~T. Walters, ``{Universality of
  Long-Distance AdS Physics from the CFT Bootstrap},''
\href{http://arxiv.org/abs/1403.6829}{{\ttfamily arXiv:1403.6829 [hep-th]}}.

\bibitem{Nakayama:2014lva}
Y.~Nakayama and T.~Ohtsuki, ``{Approaching conformal window of $O(n)\times
  O(m)$ symmetric Landau-Ginzburg models from conformal bootstrap},''
\href{http://arxiv.org/abs/1404.0489}{{\ttfamily arXiv:1404.0489 [hep-th]}}.

\bibitem{Beem:2014kka}
C.~Beem, L.~Rastelli, and B.~C. van Rees, ``{W Symmetry in six dimensions},''
\href{http://arxiv.org/abs/1404.1079}{{\ttfamily arXiv:1404.1079 [hep-th]}}.

\bibitem{Dolan:2001tt}
F.~Dolan and H.~Osborn, ``{Superconformal symmetry, correlation functions and
  the operator product expansion},''
  \href{http://dx.doi.org/10.1016/S0550-3213(02)00096-2}{{\em Nucl.Phys.}
  {\bfseries B629} (2002) 3--73},
\href{http://arxiv.org/abs/hep-th/0112251}{{\ttfamily arXiv:hep-th/0112251
  [hep-th]}}.

\bibitem{Dolan:2004iy}
F.~Dolan and H.~Osborn, ``{Conformal partial wave expansions for N=4 chiral
  four point functions},''
  \href{http://dx.doi.org/10.1016/j.aop.2005.07.005}{{\em Annals Phys.}
  {\bfseries 321} (2006) 581--626},
\href{http://arxiv.org/abs/hep-th/0412335}{{\ttfamily arXiv:hep-th/0412335
  [hep-th]}}.

\bibitem{Dolan:2004mu}
F.~A. Dolan, L.~Gallot, and E.~Sokatchev, ``{On four-point functions of 1/2-BPS
  operators in general dimensions},''
  \href{http://dx.doi.org/10.1088/1126-6708/2004/09/056}{{\em JHEP} {\bfseries
  0409} (2004) 056},
\href{http://arxiv.org/abs/hep-th/0405180}{{\ttfamily arXiv:hep-th/0405180
  [hep-th]}}.

\bibitem{Fortin:2011nq}
J.-F. Fortin, K.~Intriligator, and A.~Stergiou, ``{Current OPEs in
  Superconformal Theories},''
  \href{http://dx.doi.org/10.1007/JHEP09(2011)071}{{\em JHEP} {\bfseries 1109}
  (2011) 071},
\href{http://arxiv.org/abs/1107.1721}{{\ttfamily arXiv:1107.1721 [hep-th]}}.

\bibitem{Fitzpatrick:2014oza}
A.~L. Fitzpatrick, J.~Kaplan, Z.~U. Khandker, D.~Li, D.~Poland, and
  D.~Simmons-Duffin, ``{Covariant Approaches to Superconformal Blocks},''
\href{http://arxiv.org/abs/1402.1167}{{\ttfamily arXiv:1402.1167 [hep-th]}}.

\bibitem{Ferrara:1972xe}
S.~Ferrara and G.~Parisi, ``{Conformal covariant correlation functions},''
\href{http://dx.doi.org/10.1016/0550-3213(72)90480-4}{{\em Nucl.Phys.}
  {\bfseries B42} (1972) 281--290}.

\bibitem{Ferrara:1972ay}
S.~Ferrara, A.~Grillo, and G.~Parisi, ``{Nonequivalence between conformal
  covariant wilson expansion in euclidean and minkowski space},''
{\em Lett.Nuovo Cim.} {\bfseries 5S2} (1972) 147--151.

\bibitem{Ferrara:1972uq}
S.~Ferrara, A.~Grillo, G.~Parisi, and R.~Gatto, ``{The shadow operator
  formalism for conformal algebra. vacuum expectation values and operator
  products},''
{\em Lett.Nuovo Cim.} {\bfseries 4S2} (1972) 115--120.

\bibitem{Ferrara:1973vz}
S.~Ferrara, A.~F. Grillo, G.~Parisi, and R.~Gatto, ``{Covariant expansion of
  the conformal four-point function},''
\href{http://dx.doi.org/10.1016/0550-3213(72)90587-1}{{\em Nucl. Phys.}
  {\bfseries B49} (1972) 77--98}.

\bibitem{SimmonsDuffin:2012uy}
D.~Simmons-Duffin, ``{Projectors, Shadows, and Conformal Blocks},''
\href{http://arxiv.org/abs/1204.3894}{{\ttfamily arXiv:1204.3894 [hep-th]}}.

\bibitem{Goldberger:2011yp}
W.~D. Goldberger, W.~Skiba, and M.~Son, ``{Superembedding Methods for 4d N=1
  SCFTs},'' \href{http://dx.doi.org/10.1103/PhysRevD.86.025019}{{\em Phys.Rev.}
  {\bfseries D86} (2012) 025019},
\href{http://arxiv.org/abs/1112.0325}{{\ttfamily arXiv:1112.0325 [hep-th]}}.

\bibitem{Goldberger:2012xb}
W.~D. Goldberger, Z.~U. Khandker, D.~Li, and W.~Skiba, ``{Superembedding
  Methods for Current Superfields},''
  \href{http://dx.doi.org/10.1103/PhysRevD.88.125010}{{\em Phys.Rev.}
  {\bfseries D88} (2013) 125010},
\href{http://arxiv.org/abs/1211.3713}{{\ttfamily arXiv:1211.3713 [hep-th]}}.

\bibitem{Khandker:2012pa}
Z.~U. Khandker and D.~Li, ``{Superembedding Formalism and Supertwistors},''
\href{http://arxiv.org/abs/1212.0242}{{\ttfamily arXiv:1212.0242 [hep-th]}}.

\bibitem{Siegel:1992ic}
W.~Siegel, ``{Green-Schwarz formulation of selfdual superstring},''
  \href{http://dx.doi.org/10.1103/PhysRevD.47.2512}{{\em Phys.Rev.} {\bfseries
  D47} (1993) 2512--2516},
\href{http://arxiv.org/abs/hep-th/9210008}{{\ttfamily arXiv:hep-th/9210008
  [hep-th]}}.

\bibitem{Siegel:1994cc}
W.~Siegel, ``{Supermulti - instantons in conformal chiral superspace},''
  \href{http://dx.doi.org/10.1103/PhysRevD.52.1042}{{\em Phys.Rev.} {\bfseries
  D52} (1995) 1042--1050},
\href{http://arxiv.org/abs/hep-th/9412011}{{\ttfamily arXiv:hep-th/9412011
  [hep-th]}}.

\bibitem{Siegel:2010yd}
W.~Siegel, ``{AdS/CFT in superspace},''
\href{http://arxiv.org/abs/1005.2317}{{\ttfamily arXiv:1005.2317 [hep-th]}}.

\bibitem{Siegel:2012di}
W.~Siegel, ``{Embedding versus 6D twistors},''
\href{http://arxiv.org/abs/1204.5679}{{\ttfamily arXiv:1204.5679 [hep-th]}}.

\bibitem{Kuzenko:2006mv}
S.~M. Kuzenko, ``{On compactified harmonic/projective superspace, 5-D
  superconformal theories, and all that},''
  \href{http://dx.doi.org/10.1016/j.nuclphysb.2006.03.019}{{\em Nucl.Phys.}
  {\bfseries B745} (2006) 176--207},
\href{http://arxiv.org/abs/hep-th/0601177}{{\ttfamily arXiv:hep-th/0601177
  [hep-th]}}.

\bibitem{Kuzenko:2012tb}
S.~M. Kuzenko, ``{Conformally compactified Minkowski superspaces revisited},''
  \href{http://dx.doi.org/10.1007/JHEP10(2012)135}{{\em JHEP} {\bfseries 1210}
  (2012) 135},
\href{http://arxiv.org/abs/1206.3940}{{\ttfamily arXiv:1206.3940 [hep-th]}}.

\bibitem{Maio:2012tx}
M.~Maio, ``{Superembedding methods for 4d N-extended SCFTs},''
  \href{http://dx.doi.org/10.1016/j.nuclphysb.2012.06.011}{{\em Nucl.Phys.}
  {\bfseries B864} (2012) 141--166},
\href{http://arxiv.org/abs/1205.0389}{{\ttfamily arXiv:1205.0389 [hep-th]}}.

\bibitem{Dirac:1936fq}
P.~A. Dirac, ``{Wave equations in conformal space},''
\href{http://dx.doi.org/10.2307/1968455}{{\em Annals Math.} {\bfseries 37}
  (1936) 429--442}.

\bibitem{Mack:1969rr}
G.~Mack and A.~Salam, ``{Finite component field representations of the
  conformal group},''
\href{http://dx.doi.org/10.1016/0003-4916(69)90278-4}{{\em Ann. Phys.}
  {\bfseries 53} (1969) 174--202}.

\bibitem{Boulware:1970ty}
D.~Boulware, L.~Brown, and R.~Peccei, ``{Deep-inelastic electroproduction and
  conformal symmetry},''
\href{http://dx.doi.org/10.1103/PhysRevD.2.293}{{\em Phys.Rev.} {\bfseries D2}
  (1970) 293--298}.

\bibitem{Ferrara:1973eg}
S.~Ferrara, R.~Gatto, and A.~Grillo, ``{Conformal algebra in space-time and
  operator product expansion},''
\href{http://dx.doi.org/10.1007/BFb0111104}{{\em Springer Tracts Mod.Phys.}
  {\bfseries 67} (1973) 1--64}.

\bibitem{Weinberg:2010fx}
S.~Weinberg, ``{Six-dimensional Methods for Four-dimensional Conformal Field
  Theories},'' \href{http://dx.doi.org/10.1103/PhysRevD.82.045031}{{\em
  Phys.Rev.} {\bfseries D82} (2010) 045031},
\href{http://arxiv.org/abs/1006.3480}{{\ttfamily arXiv:1006.3480 [hep-th]}}.

\bibitem{Costa:2011mg}
M.~S. Costa, J.~Penedones, D.~Poland, and S.~Rychkov, ``{Spinning Conformal
  Correlators},'' \href{http://dx.doi.org/10.1007/JHEP11(2011)071}{{\em JHEP}
  {\bfseries 1111} (2011) 071},
\href{http://arxiv.org/abs/1107.3554}{{\ttfamily arXiv:1107.3554 [hep-th]}}.

\bibitem{Costa:2011dw}
M.~S. Costa, J.~Penedones, D.~Poland, and S.~Rychkov, ``{Spinning Conformal
  Blocks},'' \href{http://dx.doi.org/10.1007/JHEP11(2011)154}{{\em JHEP}
  {\bfseries 1111} (2011) 154},
\href{http://arxiv.org/abs/1109.6321}{{\ttfamily arXiv:1109.6321 [hep-th]}}.

\bibitem{DO1}
F.~Dolan and H.~Osborn, ``{Conformal four point functions and the operator
  product expansion},''
  \href{http://dx.doi.org/10.1016/S0550-3213(01)00013-X}{{\em Nucl.Phys.}
  {\bfseries B599} (2001) 459--496},
\href{http://arxiv.org/abs/hep-th/0011040}{{\ttfamily arXiv:hep-th/0011040
  [hep-th]}}.

\bibitem{Osborn:1998qu}
H.~Osborn, ``{N=1 superconformal symmetry in four-dimensional quantum field
  theory},'' \href{http://dx.doi.org/10.1006/aphy.1998.5893}{{\em Annals Phys.}
  {\bfseries 272} (1999) 243--294},
\href{http://arxiv.org/abs/hep-th/9808041}{{\ttfamily arXiv:hep-th/9808041
  [hep-th]}}.

\end{thebibliography}\endgroup
\bibliographystyle{utphys}

\end{document}